\documentclass[12pt]{article}

\usepackage{authblk}
\usepackage{parskip}
\usepackage[normalem]{ulem} 
\usepackage{fullpage}
\usepackage{url}
\usepackage{verbatim}
\usepackage{amstext}
\usepackage{amsmath,amssymb}
\usepackage{amsfonts}
\usepackage{mathtools}
\usepackage{float}
\usepackage{tikz}
\usetikzlibrary {arrows.meta} 
\usepackage{subcaption} 
\usepackage{color}
\usepackage{amsthm}

\tikzstyle{every node}=[circle, draw, fill=black,inner sep=0pt, minimum width=4pt]

\newcommand{\mL}{\mathbb{L}}
\newcommand{\ccL}{\mathcal{L}}

\newcommand{\link}{\prec\!\!\ast \, }

\newcommand{\mP}{\mathcal P}
\newcommand{\cP}{\mathcal P}
\newcommand{\ff}{\gamma}

\usepackage[margin=1.0in]{geometry}

\usepackage{hyperref}

\usepackage{bbold} 

\newcommand{\pkkp}{p^{(kk')}}

\newcommand{\eki}{e_{i}^{(k)}}
\newcommand{\ekpj}{e_{j}^{(k')}}

\newcommand{\Lkokt}{\ccL^{(k_1k_2)}}
\newcommand{\ffko}{\ff_{k_1}}
\newcommand{\ffkt}{\ff_{k_2}}

\usepackage{graphicx}

\newcommand{\OKR}{\Omega_{\mathrm{KR}}}

\newcommand{\SBDG}{S_{\mathrm{BDG}}}

\newcommand{\bi}{\mathrm{bilayer}}
\newcommand{\sbdg}{S_{\mathrm{BDG}}}
\newcommand{\av}[1]{\langle {#1} \rangle}

\newcommand{\OnK}{\Omega_{n,K}}
\newcommand{\OnKG}{\Omega_{n,K,\Gamma}}
\newcommand{\OnKGp}{\Omega_{n,K,\Gamma,\vec p\,}}

\newcommand{\chain}{\mathbf c}
\newcommand{\achain}{\mathbf a}
\newcommand{\NJa}{N_J^{(\alpha)}}
\newcommand{\RJa}{R^{(\alpha)}}
\newcommand{\NJakkp}{N_J^{(\alpha,k,k')}}
\newcommand{\RJakkp}{R^{(\alpha,k,k')}}
\newcommand{\NJakkpn}{N_J^{(\alpha,k,k',\vec n)}}
\newcommand{\RJakkpn}{R^{(\alpha,k,k',\vec n)}}
\newcommand{\cJa} {c^{(\alpha)}}
\newcommand{\cJakkp}{c^{(\alpha,k,k')}}
\def\bit{\begin{itemize}}
\def\eit{\end{itemize}}
\usepackage{enumerate}
 \date{}
\title{The Einstein-Hilbert Action \\ for Entropically Dominant Causal Sets}
\author{Peter Carlip${}^a \footnote{\it email:
    peter.carlip@gmail.com}$, Steve Carlip${}^b$\footnote{\it email:
    carlip@physics.ucdavis.edu}, Sumati Surya${}^c$ \footnote{\it
    email: ssurya@rri.res.in } \\
${}^a$ {\small\it Yale Department of Laboratory Medicine, 330 Cedar Street, New Haven, CT 06520, USA,}
  ${}^b${\small\it Department of Physics,}
       {\small\it University of California, }
       {\small\it Davis, CA 95616, USA}\\
   ${}^c${\small\it  Raman Research Institute, } {\small \it CV Raman Ave, Sadashivanagar, }  {\small \it  Bangalore, 560080, India} \\
}
\begin{document}

    \maketitle
    
\begin{abstract}
{In the path integral formulation of causal set quantum gravity, the quantum partition function is a phase-weighted sum 
over locally finite partially ordered  sets, which are viewed as discrete quantum spacetimes.}   It is known, however, that the number 
of ``layered'' sets---a class of causal sets that look nothing like spacetime manifolds---grows 
 superexponentially  with the cardinality $n$, giving an entropic contribution that can potentially dominate that 
 of the action.   We show here that in any dimension, the discrete Einstein-Hilbert action for a typical 
 $K$-layered causal set reduces to the simple link action to leading order in $n$.    Combined with earlier work, this 
 completes the proof that the layered sets, although entropically dominant, are very strongly suppressed in the 
 path sum of  causal set quantum gravity  {whenever the discreteness scale is greater than or equal to a  (mildly
   dimension-dependent) 
   order one multiple of  the Planck
   scale}.
 \end{abstract}
 
\section{Introduction}

A basic requirement  for any  viable theory of  quantum gravity is  that classical spacetime must emerge in the small $\hbar$
regime. In the language of path integrals, this means that the saddle point contributions to the partition function come from 
classical spacetimes. In the causal set  approach to quantum gravity \cite{blms}, the path integral over continuum spacetimes is
replaced by a path sum over causal sets, that is, locally finite partially ordered sets. For a fixed cardinality $n$, the partition 
function is the path sum 
\begin{equation}
  Z(\Omega_n)=\sum_{C \in \Omega_n} \exp{\frac{i}{\hbar} S(C)} ,
  \label{part.eq} 
\end{equation} 
where $\Omega_n$ is the space of all $n$-element causal sets and $S(C)$  is a causal set action. In this discrete
setting,  continuum spacetimes are replaced by causal sets obtained from the continuum via a random
discretisation. The question is whether such  ``continuumlike''  causal sets dominate the causal set partition function in
the large $n$ limit.  

A potential challenge to the dynamical emergence of the continuum comes from a
class of ``layered'' causal sets whose entropic contribution grows  superexponentially with 
cardinality $n$ \cite{kr,dharone,pst}.  These causal sets do not have a continuum approximation{, by which we mean} they cannot be 
obtained as random discretisations of any continuum spacetime. It is therefore important to know  whether any 
reasonable choice of action $S(C)$ is capable of suppressing  this entropic contribution in the small $\hbar$ limit.

Nonclassicality in the context of the path integral  is not per se surprising.  Even for a free particle, the
space of paths is dominated by those that are highly nonclassical and only $C^0$ \cite{Morette}.
 As we know from textbook arguments, though, these paths interfere destructively, allowing the
classical paths to dominates in the $\hbar \rightarrow 0$ limit. Demonstrating such behavior  in quantum gravity is
of course more challenging, but progress has been made in recent years in causal set theory, where it has been
shown that the path integral contributions from two- and three-layered causal sets  are  suppressed at leading order 
\cite{lc,ams,ccs}.   Our present work adds a finishing touch, showing that the contribution from all $K$-layered 
causal sets with $K\ll n$  is suppressed whenever the discreteness scale is greater than or equal to a (mildly dimension-dependent)
  order one multiple of the Planck scale.

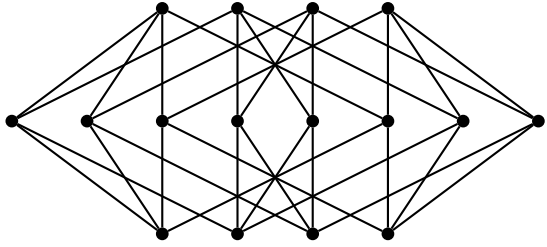
\begin{figure}[h]
\begin{center}
\begin{tikzpicture} [style=thick] 
  \node (a1) at (0,0){};
  \node (a2) at (1,0) {};
  \node (a3) at (2,0) {};
  \node (a4) at (3,0) {};
  
  \node (a5) at (-2,1.5) {};
  \node (a6) at (-1,1.5) {};
  \node (a7) at (0,1.5) {};
  \node (a8) at (1,1.5) {};
  \node (a9) at (2,1.5) {};
  \node (a10) at (3,1.5) {};
\node (a11) at (4,1.5) {}; 
\node (a12) at (5,1.5) {};

 \node (a13) at (0,3) {};
 \node (a14) at (1,3) {}; 
 \node (a15) at (2,3) {}; 
 \node (a16) at (3,3)   {};

  \draw (a1) -- (a5);
  \draw (a1) -- (a6);
  \draw (a1) -- (a7);  
  \draw (a1) -- (a10);
  \draw (a2) -- (a5);
  \draw (a2) -- (a8);
  \draw (a2) -- (a9);
  \draw (a2) -- (a11);  
  \draw (a3) -- (a6);  
  \draw (a3) -- (a8);
  \draw (a3) -- (a9);
  \draw (a3) -- (a12);
  \draw (a4) -- (a7);  
  \draw (a4) -- (a10);
  \draw (a4) -- (a11);
  \draw (a4) -- (a12);
 \draw (a13) -- (a5);
  \draw (a13) -- (a6);
  \draw (a13) -- (a7);  
  \draw (a13) -- (a10);
  \draw (a14) -- (a5);
  \draw (a14) -- (a8);
  \draw (a14) -- (a9);
  \draw (a14) -- (a11);  
  \draw (a15) -- (a6);  
  \draw (a15) -- (a8);
  \draw (a15) -- (a9);
  \draw (a15) -- (a12);
  \draw (a16) -- (a7);  
  \draw (a16) -- (a10);
  \draw (a16) -- (a11);
  \draw (a16) -- (a12);
 \end{tikzpicture}
\caption{The Hasse diagram of a KR order. The diagram only shows links, that is, nearest neighbors; the remaining
relations follow from transitive closure.}
\label{kr.fig}
\end{center}
\end{figure}

The combinatorial results on the asymptotic growth in the number of  layered partially ordered sets
predate causal set theory.  As shown in \cite{kr,dharone,pst}, the  leading contribution to $\Omega_n$ comes from the three-layered
Kleitman-Rothschild (KR) sets, with\footnote{The notation $|\Omega|$ means the cardinality of the set $\Omega$, that
is, the number of elements in $\Omega$.} $|\Omega_{KR}| \sim
2^{\frac{n^2}{4}+ \frac{3n}{2}   +o(n)}$ \cite{kr} (see  Fig.\ \ref{kr.fig}). The next 
largest contribution comes from the {symmetric} bilayer sets, with $|\Omega_{\bi}| \sim 2^{\frac{n^2}{4}+O(n)}$.
This is is the start of  a hierarchy of subdominant contributions, the next in line being the four-layer sets, then the 
five-layer  sets, and so on \cite{dharone, pst}.  In general, for $K \ll n$,
\begin{equation}
  |\Omega_{K}| \sim 2^{{c(K) n^2}+o(n^2)},
\label{ckdef.eq}
\end{equation} 
where $c(K) \leq \frac{1}{4}$ is a weakly monotonically decreasing function for $K \geq 4$ \cite{pst}.
 
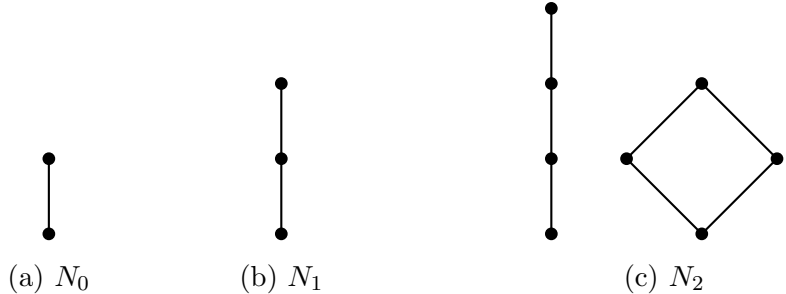
\begin{figure}[!bhtp]
  \hspace{1cm} 
\begin{subfigure}[b]{.3\linewidth}  
\begin{center}
\begin{tikzpicture}[style=thick] 
  \draw (-2,0) node {};
  \draw (-2,1) node {};
\draw (-2,0) --(-2,1); 
\end{tikzpicture}
\caption{$N_0$}
\end{center}
\end{subfigure}\hspace{-2cm} 
\begin{subfigure}[b]{.3\linewidth}  
\begin{center}
\begin{tikzpicture}[style=thick] 
\draw (-2,0) node {};
\draw (-2,1) node {};
\draw (-2,2) node {};
  \draw (-2,0) -- (-2,1);
  \draw (-2,1) -- (-2,2);
\end{tikzpicture}
\caption{$N_1$}\end{center}
\end{subfigure}
\begin{subfigure}[b]{.3\linewidth}  
\begin{center}
\begin{tikzpicture}[style=thick] 
 \draw (-2,0) node {};
  \draw (-2,1) node {};
  \draw (-2,2) node {};
   \draw (-2,3) node {};
  \draw (-2,0) -- (-2,1);
  \draw (-2,1) -- (-2,2);
  \draw (-2,2) -- (-2,3);
  \draw (0,0) -- (-1,1);
   \draw (0,0) -- (1,1);
\draw (-1,1) -- (0,2);
\draw (1,1) -- (0,2);
\draw (0,0) node {}; 
 \draw (-1,1) node {};
  \draw (1,1) node {};
  \draw (0,2) node {};
\end{tikzpicture}
\caption{$N_2$}
\end{center}
\end{subfigure}
\caption{Examples of intervals of size $0,1$ and $2$:  $N_0$ counts the number of type  (a) intervals or links in $C$, $N_1$ the
  number of type  (b)  intervals,  and $N_2$ the number of any one of the two type (c)  intervals. In diagram (c) the
totally ordered  ``chain'' has exactly  one $4$-element path and the ``diamond'' causal set to its  right has exactly  two  $3$-element paths.} 
\label{ints.fig} 
\end{figure}

It is, of course, possible to concoct an action that would suppress this overwhelming entropic contribution to
$Z(\Omega_n)$, but such an action should have a reasonable physical motivation.  A natural choice is  the 
Benincasa-Dowker-Glaser (BDG)  action \cite{bd,dg,glaser}, the discrete version of the Einstein-Hilbert action
in $d$ spacetime dimensions.  For a causal set $C$ containing $n$ elements, this action is
\begin{equation} 
\frac{1}{\hbar} S^{(d)}_{BDG}(C) = -\alpha_d\biggl( \frac{\ell}{\ell_p} \biggr)^{d-2} \biggl( n + \frac{\beta_d}{\alpha_d}
  \sum_{J=0}^{n_d} C_J^{(d)} N_J \biggr) ,
\label{bdg.eq} 
\end{equation} 
where  $\ell_p$ is the Planck length, $\ell$ is a discreteness scale, and $\alpha_d$, $\beta_d$,  and $C_J^{(d)}$ are known
dimension-dependent constants.\footnote{See \cite{glaser} for closed form expressions for $\alpha_d,\beta_d$ and
  $C_J^{(d)}$, where their $i$ is our $J+1$.}  The 
dependence on the causal set $C$ comes through the interval abundances $N_J$, which count the number
of intervals in $C$ with exactly  $J$ elements---that is, the number of pairs  $e,e'\in C$  for which the cardinality 
$|\{e''|e \prec e'' \prec e'\}| = J$.  (We adopt the irreflexive convention $e \nprec e$.) Thus,  $N_0$ counts the 
number of $0$-element intervals or
\emph{links} (nearest neighbors), $N_1$ the number of $1$-element intervals (next-to-nearest neighbors),
and so on. Fig.\ \ref{ints.fig} shows the first few $N_J$. For causal sets that are approximated by continuum 
spacetimes, it has been shown that this  discrete action goes over to  the Einstein-Hilbert action (up to boundary 
terms) in the  continuum limit \cite{bd,dg,glaser}.
Here, note that the  spacetime dimension $d$ is not intrinsic to the causal set, but imposed by hand.  We will
see in what follows that this dependence on $d$ is only marginally relevant to the question at hand. 

 The first concrete step towards answering the entropy versus action question in causal set theory was taken in 
\cite{lc}, where the contribution to $Z(\Omega_n)$ of  the class of bilayer causal sets $\Omega_\bi$ was
analyzed.  These are the next subdominant class of partially ordered sets in $\Omega_n$ after $\OKR$.  For 
bilayer sets, $\sbdg$ simplifies drastically,  since  the only  intervals are links, 
 which  contribute only to $N_0$. Thus,
$N_J =0$ for all $J>0$, and $\sbdg$ reduces to the ``link action'' 
\begin{equation} 
\frac{1}{\hbar} S^{(d)}_{\mathrm{link}}(C) = -\alpha_d\biggl( \frac{\ell}{\ell_p} \biggr)^{d-2} \biggl( n + \frac{\beta_d}{\alpha_d}
N_0 \biggr), 
\label{linkaction} 
\end{equation}
where we have used the fact that in Eqn.\ (\ref{bdg.eq}), the coefficient $C_0^{(d)}=1$ for all  $d$ \cite{glaser}.  
The maximum number of links in a bilayer set is $\frac{n^2}{4}$ (from $\frac{n}{2}$ elements in layer $1$ each linked to
$\frac{n}{2}$ elements in layer $2$).  For a general bilayer set, we can thus write $N_0=\frac{1}{4} p n^2$, with a
``linking fraction'' $p \in[0,1]$.  In \cite{lc}, a bounding argument was used to count the number of causal sets for a given 
$p$, and the path integral was approximated as an integral over $p$ and evaluated by the method of steepest
descent.  The result was that the bilayer contribution to the partition function was superexponentially suppressed
whenever\footnote{{This makes a small correction to  an estimate from  previous
    work \cite{lc} where the $\beta_d$ dependence  was mistakenly dropped.}}%
\begin{align}
  {\tan\left[\frac{1}{2}\beta_d\left(\frac{\ell}{\ell_p}\right)^{d-2}\right] > \left(\frac{27}{4}e^{-1/2}-1\right)^{1/2}.}
\label{l2}
\end{align}
This will be satisfied whenever $\ell>\ell_0$, where the lower limit
   $\ell_0$ is mildly dimension dependent but always of order $\ell_p$.
   For $d=4$, $\ell_0\approx1.136\ell_p$, and  for general $d$, 
    $1.13  < (\ell_0/\ell_p)  <  2.33 $, giving a very conservative constraint on
   the discreteness scale.

For a   $(K>2)$-layer partially ordered set, however,  the $N_J$ need not be zero for $J>0$. For a similar strategy to work  
one would have to find the number of causal sets that contribute to each value of the BDG action,  a considerably more 
complicated calculation.  As a warm-up,  \cite{ams} considered the path sum with the BDG action replaced by the
simpler link action \eqref{linkaction}.  The link action is determined by the linking fraction $p$, and it is again possible, 
to leading order in $n$, to find the  number of causal sets as a function of $p$.  As in the bilayer case, the link action 
was shown to suppress all $K$-layer partially ordered sets with $K\ll n$, subject to the same condition  \eqref{l2} on the
discreteness scale.

More recently we showed that, to  leading order,  the BDG action for KR orders \emph{is} the link action \cite{ccs}. 
Since the 
arguments of \cite{lc,ams} only need the leading order contribution, this suffices to show that KR orders are in fact suppressed 
by the full BDG action.  In this work we extend this analysis  to  all $K$-layer causal 
sets with $K\ll n$. This completes the proof that the entropic contribution
from the $K$-layer causal sets is strongly suppressed by the BDG action in the causal set path sum. 
Because the link term appears in the BDG action in all dimensions, this suppression is dimension independent. 

In section \ref{prelim.sec} we start with some definitions and give a brief review of the $K=3$ case of \cite{ccs}.  We
use a slightly different perspective, which is easier to generalise to $K>3$.  Section \ref{main.sec} contains our main
results. We begin by outlining our strategy  in section \ref{strategy.ssec}.  In section \ref{mainK4.ssec} we implement
this explicitly for the  $K=4$ case for intervals of size $J=1,2,3$, since this gives us some 
important general  insights into the leading order
contributions to $N_J$. In section \ref{mainKJ.ssec} we prove our main
result, namely that for a typical $K$-layer partially ordered set with $K\ll n$, the $N_J$ for $J>0$ contribute negligibly to action for large $n$.  In section
\ref{conclusions.sec} we discuss the implications of our results and some of the remaining open questions. 

\section{Preliminaries} 
\label{prelim.sec} 

We start with the basic definition.
A \emph{labelled $n$-element  causal set} $C$ is a locally
finite partially ordered set, on $n$-elements $\{e_1, e_2 \ldots e_n\}$, where the order relation $\prec$ is taken to be irreflexive,
i.e.,  $e_i \not \prec e_i$.   
As in any partially ordered set, the order relation is \emph{transitive}---that is, $e_i \prec e_j, e_j \prec e_k \Rightarrow e_i \prec e_k$.
The irreflexive condition thus implies that there are no elements $e_i,e_j\in C$ for which both $e_i\prec e_j$ and $e_j\prec e_i$.
 \emph{Local finiteness} is the statement that  $\forall e_i,e_j \in C$ there are at most a finite number of $e_k\in C$ 
for which  $e_i \prec e_k \prec e_j$; this is a discreteness condition.

 The distinguishing label  for each element $e_i$ in the causal
 set $C$  is,  in  essence,  a choice of discrete coordinates.  Labels can simplify calculations, but ultimately they, like coordinates, must be ``gauged
away.''   Any physical results  therefore should be independent of labels. 
The  permutation  group $S_n$ on $n$ elements acts on the set of $n$-element labelled causal sets  as a 
gauge group, with each orbit corresponding to a single
\emph{unlabelled} 
causal set $[C]$ of all possible relabellings of a labelled causal set $C$.  A typical orbit has $n!$ causal sets,  
but there are exceptions. If a labelled causal set has two elements $e_i,e_j$ with identical
pasts and futures, then permuting the labels of $e_i$ and $e_j$ ($i
\leftrightarrow j)$  leaves the labelled  causal set
unchanged, so the orbit on which it lies  has
fewer than $n!$  labelled causal sets.  An exact factorization of the
gauge group thus becomes complicated.

In this paper we will use the set $\Omega_n$ of  labelled $n$-element causal sets  since
this greatly simplifies the calculations.  However, since we are interested
only in the leading order contributions, our results also hold for
unlabelled causal sets and are therefore gauge independent.  The reason is the following. The  worst
possible overcounting of causal sets  due to labelling is by a factor
of $n!$, which asymptotically goes as $\sim 2^{n \log_2 n}$.  Such a
factor is subdominant with respect to the leading order
contributions, as we will see.

We now construct various sub-causal sets  of  $C$. 

An \emph{$r$-element chain}  in $C$ is a totally ordered subset $e_{i_1} \prec e_{i_2} \ldots \prec e_{i_r}$ of $r$ elements, and will be said
  to be of \emph{length} $r$. Thus, a relation $e_i
\prec e_j$ is a $2$-element chain and is of length $2$. The \emph{height} $h(C) $ of the causal set is the length of the longest  chain
in $C$.   An \emph{antichain} is a set of unrelated elements, that is, a set $A$ such that
$e_i\not\prec e_j$ for any $e_i,e_j\in A$.

For any  $e_i\prec e_j$, the \emph{interval} ${(e_i,e_j)}$  is the
set $\{e_k\in C |e_i\prec e_k\prec e_j\}$, examples of which are shown in Fig.\ \ref{ints.fig}.   An interval is the
discrete,  
causal set analog of a causal diamond, or Alexandrov interval $I(p,q) \equiv I^+(p) \cap I^-(q)$. As in the continuum,
${(e_i,e_j)}$ is causally convex, i.e.,  for any pair $e_k, e_r
\in e_i \cup e_j \cup {(e_i,e_j)}$, ${(e_k,e_r) \subseteq
  (e_i,e_j)}$. Hence for every $e_i \prec e_j$ we can, in a consistent way,  define
the \emph{height}  $h(e_i,e_j)$ as the length of the longest chain from $e_i$ to $e_j$, which is also the height of the
subcausal set  $e_i\cup e_j\cup {(e_i,e_j)}$.   The \emph{size} of an interval ${(e_i,e_j)}$ is its cardinality $|{(e_i,e_j)}|$ (our convention being  \emph{not}  to include
the endpoints $e_i,e_j$ in the counting).   The quantities $N_J$ associated with a causal set $C$,  which  appear in the BDG
action Eqn.\ \eqref{bdg.eq}, count the number of $J$-size or $J$-element intervals in $C$.  

A pair of elements $e_i,e_j \in C $ is said to be  \emph{linked}  if $e_i \prec e_j$ and there exists  no $e_k$ such that $e_i \prec
e_k \prec e_j$. Thus $e_i$ and $ e_j$ are ``nearest neighbors.''  We
denote this {\it link} relationship by $e_i \link
e_j$. Whenever  $e_i \link
e_j$, $|{(e_i,e_j)}|=0$, and hence links contribute to the $0$-element intervals $N_0$ in the BDG action
Eqn.\ \eqref{bdg.eq}.

All the relations ({$e \prec e'$}) in $C$ are   uniquely determined from the links by  implementing transitive
closure. $C$ can therefore be  uniquely represented by its Hasse diagram, as in Fig.\ \ref{kr.fig}, or equivalently by the $n \times n$
\emph {link matrix} $\ccL$   where 
\begin{equation}
  \ccL_{ij}=
  \begin{cases}
    1 & e_i \link e_j \\
0 & {\text{ otherwise}}.\\ 
\end{cases}
\label{linkmatrix.eq} 
\end{equation}
Thus, the total number of links in $C$ is 
\begin{equation}
N_0 = \sum_{i=1}^n \sum_{j=1}^n \ccL_{ij}. 
  \end{equation} 
 One can extract more information from higher powers of $\ccL$, as we shall soon see.

  Consider an  $r$-element chain where every relation in the chain is a link. We will refer to this as an   
  \emph{$r$-element path}  of
   length $r$,  i.e., the totally ordered subset  $\{e_{i_1}, e_ {i_2}, \ldots e_{i_r} \}$ with $e_{i_1} \link
   e_{i_2}\ldots  \link e_{i_r}$.  Thus, a  linked pair $e_i \link e_j$ is a  $2$-element path of length $2$.  Since the $\link$ relation is 
a nearest neighbor relation, unlike chains,  there can be at most one $2$-element path between two
elements. This is reflected by the fact that the matrix elements of $\ccL$ are either $0$ or $1$.

However, there can be multiple $r$-element paths from $e_i$ to $e_j$ when $e_i \prec e_j$,  for $r>2$. Consider the squared matrix $\ccL^2$
  \begin{equation}
 ( \ccL^2)_{ij}= \sum_{k=1}^n \ccL_{ik} \ccL_{kj}.
\end{equation}
The sum picks up a $1$ for each  $ e_{k}  $ for which $  e_i \link e_{k} \link e_j$ ,  and a $0$
otherwise. Hence $(\ccL^2)_{ij}$ is equal to  the number of  $3$-element paths from $e_i$ to $e_j$.  
In general, for  $r \geq 2$, $(\ccL^{r-1})_{ij}$ is the number of $r$-element paths from $e_i$ to $e_j$.  It is important
to note that  $\ccL^{r-1}$  does not  directly give information about  $N_J$, for $J>0$. This is illustrated in 
Fig.\ \ref{L32.fig}.  However, one can use powers of $\ccL$ to put bounds on the $N_J$.

In the course of this work we will be counting the number of labelled causal
sets in various subclasses $\Omega'_n \subset
\Omega_n$ characterized by a variety of properties. We define the probability that a causal set has some property $P$  by the {counting measure} 
\begin{equation}
\mathcal P(C\in {\Omega'_n}\ \hbox{has property $P$}) = 
\frac{\hbox{number of sets in ${\Omega_{n}'}$ with property $P$}}{\hbox{total number of sets in ${\Omega_{n}'}$}}  .
\label{prob.eq} 
\end{equation}

 We are now in a position to characterise  the class of causal sets of interest to us in this paper. 

Following \cite{pst},  a causal set $C \in \Omega_n$ is said to be \emph{$K$-layered} if it  can be partitioned into $K$ disjoint antichains $\mL_i$,  
$C= \mL_1 \cup \mL_2 \cup \ldots \mL_K$,  such that:
\begin{itemize}
\item If $e\prec e'$ with   $e \in \mL_k$ and $  e' \in \mL_{k'}$, 
  then  $k<k'$.
  \item   For every  $e \in \mL_k$ and $  e' \in \mL_{k'}$ with
    $k'>k+1$, $e \prec e'$. 
  \end{itemize} 

We denote the  set of all $K$-layered partially ordered sets in
$\Omega_n$ by $\OnK$, which, according to our definition,\footnote{{For an example of a causal set  which
  cannot be assigned  layers for {\it {any}}  choice of $K$, see \cite{ams}.}}  is a strict
subset of $ \Omega_n$.

For $C \in \OnK$, let $\ff_kn \equiv |\mL_k| $, the number of
elements in layer $k$, where, since every element is in exactly one layer, $\sum_{k=1}^K \gamma_k=1$. 
We define the \emph{linking fraction} $p^{(k(k+1))}$ between consecutive layers $\mL_k$ and $\mL_{k+1}$ as
    the ratio of the total number of links between these layers in $C$
    to the maximum number of ``possible'' links  $\ff_k\ff_{k+1}n^2$
    between the two layers.

We pause here to note that the assignment of layers to elements of a causal set,
even when possible,  may not be unique.    For example, in a $K=2$
layer causal set,  an element can be 
 unrelated to all others, and hence assigned either to $\mL_1$ or
 $\mL_2$. For $K =3$, however, such an element can only be assigned
 to $\mL_2$, while for $K>3$ there can be no such elements. In
 general, if an element $e \in \mL_k,\,  k>1$, then it can be assigned to
 $\mL_{k-1}$ only if  (a) $\not\!\exists \, e' \in \mL_{k-1} $ such that $e'\prec
 e$ and (b) $\forall \, e'' \in \mL_{k+1} $,  $e \prec
 e''$. A similar ``time-reversed''  pair of conditions need to be
 satisfied in order for $e \in \mL_k, \, k<K$ to be assigned to $\mL_{k+1}$.  Clearly,
 there can be no further ambiguity because of the second condition in
 the definition of the $K$-layer orders. This nonuniqueness means that the same set can appear more than once in a
list of layered sets. Such double counting is very rare, but can be completely
eliminated by requiring that each element be assigned the earliest layer in
which it can appear.

We consider a further subdivision of  $\OnK$. Define 
 $\OnKG \subset \OnK$ to be the set of $n$-element $K$-layer causal sets with fixed 
\emph{filling fraction} $\Gamma_K=(\gamma_1, \gamma_2, \ldots \gamma_K)$.  It is convenient to use the
ordering of the layers $k = \{1, \ldots K \}$ to label the elements of $C \in {\OnKG}$.  We denote the elements of
$\mL_k$ by $\eki$, where the upper bracketed index labels the layer and the lower index $i\in [1, \ff_k n]$ distinguishes elements
within that layer.  The ordering of the layers allows the link matrix to be written in block
diagonal form
\begin{equation} \ccL = 
\begin{pmatrix}
0 & \ccL^{(12)}  & \ccL^{(13)} & \ldots & \ccL^{(1K)} \\
0 & 0  & \ccL^{(23)} & \ldots & \ccL^{(2K)} \\
0 & 0 & \ldots & \ldots & \ldots \\
0 & 0 & \ldots   & 0 & \ccL^{((K-1)K)} \\
0 & 0  & \ldots   & 0 & 0\\ 
\end{pmatrix}, 
\end{equation}
where each $\Lkokt$ is itself the  $\ffko n \times \ffkt n$ link matrix 
\begin{equation}
  \Lkokt_{ij}=
  \begin{cases}
    1 & e_i^{(k_1)}\link e_j^{(k_2)}\\
0 & {\text{ otherwise}}\\ 
  \end{cases} 
\end{equation}
for the subcausal set $\mL_{k_1} \sqcup \mL_{k_2}$.  Henceforth, we will consider only those labelled causal sets that are
consistent with this ``layer-induced'' labeling. 

 {Let   $\vec p = \{p^{(k(k+1))}\} $ be a set of linking
fractions  between consecutive layers and let $\OnKGp \subset \OnKG$ be the associated subset of
labelled causal sets with linking fractions $\vec p$.
In this work we  will calculate, to
leading order in $n$,  the number of causal sets in $\OnKGp$ with
$N_J=R_J$ for a given $J \ll n$ and $R_J \sim n^2$.}

{ The $p^{(k(k+1))}$ can also be viewed as the probability that a
given   $e_i^{(k)} \in \mL_k$ is linked to a given $e_j^{(k+1)} \in
\mL_{k+1}$,  using the counting measure, as in Eqn.\eqref{prob.eq}.  Consider a
fixed causal set in $\OnKGp$, and now let us generate from it all other
causal sets in $\OnKGp$, with   $p^{(k(k+1))}\gamma_k \gamma_{k+1}n^2$ links between  $\mL_k$ and
$\mL_{k+1}$, i.e.,  without changing any of the links between other
consecutive layers.  The number of such causal sets is
\begin{equation} \binom{\gamma_k \gamma_{k+1}n^2}{
    p^{(k(k+1))}\gamma_k \gamma_{k+1}n^2}.
\end{equation}
Hence we can define the probability for a
given $e_i^{(k)} \in \mL_k$ to be  linked to a given $e_j^{(k+1)} \in
\mL_{k+1}$ as   
\begin{equation}
\binom{\gamma_k \gamma_{k+1}n^2-1}{
  p^{(k(k+1))}\gamma_k \gamma_{k+1}n^2-1} \binom{\gamma_k \gamma_{k+1}n^2}{
  p^{(k(k+1))}\gamma_k \gamma_{k+1}n^2}^{-1} =  p^{(k(k+1))}. 
  \end{equation} 

In general let  $\pkkp$  denote the 
    probability for an element  from $\mL_k$ to be linked
  to an element in $\mL_{k'}$ for $k< k'$. While the $p^{(k(k+1))}$ can all be chosen
independently,  this is not in general possible for $p^{(k'k)}$ when 
$k'>k+1$ because of constraints from transitivity.  For example, if
$p^{(k(k+1))}=1$ for all $k$ (i.e., every pair in  
$\mL_k$ and $\mL_{k+1}$ are maximally linked),  then transitivity
implies that $\pkkp=0$ for
all $k'>k+1$.  Conversely,  for $k'>k+1$, $\pkkp$ is proportional to the probability for there to be \emph{no} $r$-element paths from $e_i^{(k)}$ to $e_j^{(k')}$ for $2<
r \leq   k'-(k+1)+2$. This will be a crucial feature of our construction.}

The analysis  of \cite{kr,dharone,pst}  shows that the leading order contribution to $\OnK$, determined by $c(K)$ in
Eqn.\ \eqref{ckdef.eq},  comes from a  subclass of  \emph{dominant configurations} ${\mathcal Q}_K \subset
\OnK$. The causal sets in this subclass have a fixed filling fraction $\Gamma_K$ as well as a fixed {linking fraction} $p$ between consecutive
layers. The $\Gamma_K$ and $p$
for these dominant configurations themselves are functions of the   number of total relations  $\mathbf R$.  Thus, for every $K$,
the coefficient $c(K)$ in Eqn.\ \eqref{ckdef.eq} depends not only on $K$  but also on $\mathbf R$.  Our results 
apply more generally to causal sets in $\OnK$, and hence also include this dominant  subclass of configurations.

We  stated in the introduction  that causal sets in  $\OnK$ are not \emph{continuumlike}.  What we mean by this
qualification 
is the following. A causal set $C$ is said to be
continuumlike if it can be obtained from a  continuum $d$-dimensional spacetime $(M,g)$ by selecting {events}  
from $M$ randomly via a  Poisson distribution, at density $\rho$ with respect to the volume element $\sqrt{|g|}$, with causal relations inherited
from $(M,g)$. Thus, the probability for there to be $n$ causal set elements in a spacetime 
volume $V$ is given by 
\begin{equation}
P_V(n)= \frac{1}{n!} (\rho V)^n e^{-\rho V} . 
  \end{equation} 
Randomness combined with 
local finiteness makes it possible to recover, on average, the continuum spacetime volume of a region from the (finite) 
number of causal set elements in that region, since  $\av{N}=\rho V$.  The density $\rho$ 
determines a characteristic discreteness scale $\ell {=\rho^{-\frac{1}{d}}}$, where $d$ is the dimension of $M$.  Below $\ell$,
 the causal set contains little information about  $M$.  But on scales large compared to $\ell$,  the dimension and 
 much of the  topology and geometry of $(M,g)$ can be reconstructed from the causal set (see \cite{lr} for a review.)

  Consider a causal set  that is obtained
   from a Poisson sprinkling into a causally convex region of a
   spacetime at sufficiently high density $\rho$, such that $n \gg 1$.
   While such a set can be decomposed into an ordered set of
   non-overlapping 
   antichains $\{A_i\}$, it will typically {\it not} be a $K$-layer
   order. For such
     causal sets, the  probability that an 
   element in $A_i$ is related to one in $A_{j}$ depends on the
   continuum spacetime volume between the two elements in the
   embedding. This in turn depends  on where the two elements lie on
   their respective antichains. Thus, the requirement that 
   the linkage probability is the same for all pairs in $A_k$ and $ A_{k+1}$, or that all elements
     in $A_k$ are to the past of all elements in $A_{k'}$ for
     $k'>k+1$, is not consistent with the probabilities obtained
     from a sprinkling.  This is closely related to the fact that causal sets
   that are continuumlike preserve local Lorentz invariance. Thus, as long
   as $n \gg K$,  the typical $C \in \OnK$
 is \emph{not} continuumlike.

\subsection{Review of the  three-layer ($K=3$) case \label{secK3.ssec}}

\begin{figure}[h]
\begin{center}
\begin{tikzpicture}[style=thick]
  \draw[line width=1pt, dashed] (0,0) -- (4,0);
  \draw[line width=1pt, dashed] (0,1) -- (4,1);
  \draw[line width=1pt, dashed] (0,2) -- (4,2);
\draw[color=black] (2,0) -- (2,1);
\draw[color=black] (2,1) -- (2,2);
\draw[color=black] (2,0) -- (0,1);
\draw[color=black] (0,1) -- (2,2);
\draw[color=black] (2,0) -- (1,1);
\draw[color=black] (1,1) -- (2,2);
\draw[color=black] (2,0) -- (1,1);
\draw[color=black] (1,1) -- (2,2);
\draw[color=black] (2,0) -- (3,1);
\draw[color=black] (3,1) -- (2,2);
\draw[color=black] (2,0) -- (4,1);
\draw[color=black] (4,1) -- (2,2);
\draw[color=black] (2,0) node[label=below:$e_i^{(1)}$]{}; 
\draw[color=black] (2,1) node {};
\draw[color=black] (2,2) node[label=above:$e_j^{(3)}$]{};
\draw[color=black] (1,1) node {};
\draw[color=black] (0,1) node {};
\draw[color=black] (2,1) node {};
\draw[color=black] (3,1) node {};
\draw[color=black] (4,1) node {};
\end{tikzpicture}
\caption{{The number of $1$-element paths for $K=3$ from  $e_i^{(1)} \in \mL_1$ to  $e_j^{(3)} \in \mL_3$ is also the size
    of the interval ${(e_i^{(1)} , e_j^{(3)} )}$.}}
\label{3lev.fig}
\end{center}
\end{figure}
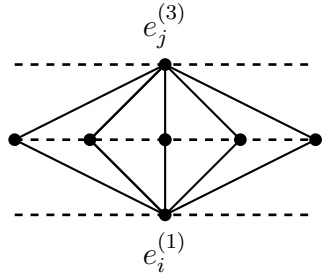 

In \cite{ccs} it was shown for $K=3$ that while $N_0 \sim n^2$, the  $N_{J>0}$ grow at most as $n$, and thus
contribute negligibly to the BDG action at large $n$.  
We review this construction briefly here to set the stage for the calculation for  $K>3$.

{The
  method used in \cite{ccs} depended on properties of the link matrix.  For $K=3$, $\ccL^{r-1}=0$ for  $r>3$,
  i.e., the maximum length of a path  is  $r=3$.  Moreover, for $e^{(1)}_i \in \mL_1$ and $e_j^{(3)}   \in \mL_3$, 
  $\ccL^2_{ij}$ not only counts the number of $3$-element
  paths from $e^{(1)}_i$ to $e_j^{(3)} $, it \emph{also} counts the {size} of the interval ${(e_i^{(1)} , e_j^{(3)} )}$. This simplifies the
  $K=3$ calculation considerably.

 While the link matrix  can still be used for $K>3$,  the calculation becomes more involved when there are more than
 three layers. For this reason, we will give a slightly different, although ultimately equivalent, proof for $K=3 $ which we
 can then  generalise  to arbitrary K. }

Let  $C \in \Omega_{n,3,\Gamma,\vec p}$ for some choice of
  $\Gamma=(\gamma_1, \gamma_2, \gamma_3)$ and consecutive linking fractions $\vec
  p$. Fix elements $e_i^{(1)} \in \mL_1$ and $e_j^{(3)}   \in \mL_3$.  As shown in Fig.\ \ref{3lev.fig}, 
  the
  interval ${(e_i^{(1)} , e_j^{(3)} )}$ is a set of causally unrelated elements---an antichain---all lying in layer $2$. Let us
first ask, for a fixed $e_i^{(1)} $ and $e_j^{(3)} $, for the probability $q(J)$ that this antichain has exactly $J$ elements, with 
$0\le J\le \gamma_2n$.  For a set of $J$ \emph{fixed} elements in layer $2$, this probability is clearly
$$(p^{(12)}p^{(23)})^J (1-p^{(12)}p^{(23)})^{\ff_2 n-J}$$
where the first factor is the probability that the chosen $J$ elements are included and the second factor is the
probability that the remaining $\ff_2 n-J$ elements are excluded.  Since there are $\binom{\ff_2n}{J}$ ways of
selecting the $J$ elements in layer $2$, the total probability that ${(e_i^{(1)} , e_j^{(3)})}$ has exactly $J$ elements is 
 \begin{equation}
  q(J)= \binom{\ff_2n}{J}(p^{(12)}p^{(23)})^J (1-p^{(12)}p^{(23)})^{\ff_2 n-J}. 
  \label{q3.eq}
\end{equation}
For $n$ large and $J\ll n$,
\begin{equation} 
\binom{\ff_2n}{J} \sim  \frac{(\ff_2n)^J}{J!} .
\end{equation} 
Defining $\alpha = 1-p^{(12)}p^{(23)} <1$, we have
\begin{equation}
  q(J) \sim \frac{(\ff_2 p^{(12)} p^{(23)})^J}{J!} n^J \alpha^{\ff_2 n} = A_J n^J 2^{-\ff_2 n|\log_2\alpha|} ,
  \label{q4.eq}
\end{equation}
which isolates the leading order $n$-dependence  for $\ff_2>0$.  Note that for $n$ large, $q(J)$ is exponentially small;
this will be the key cause of suppression of $J$-element intervals for  three-layer sets.
 
For $J>0$ we can now ask for the probability  {$\mP(N_J=R_J)$}  that the number of $J$-element intervals in the causal set $C$ is
 exactly $N_J=R_J$.  {Only intervals of the
   type $(e_i^{(1)}, e_j^{(3)})$ can contribute to $J>0$}.   There are $\ff_1\ff_3 n^2$ possible pairs $e_i^{(1)} \in \mL_1,  e_j^{(3)}\in \mL_3$,  hence  by the
 same reasoning that led to (\ref{q3.eq}),
 \begin{equation} 
  \mP(N_J=R_J) = \binom{\ff_1\ff_3 n^2}{R_J}q(J)^{R_J}(1-q(J))^{\ff_1\ff_3 n^2-R_J}. 
  \label{pnr3.eq}
\end{equation}
Since the asymptotic behavior of $q(J)$ is dominated by the term $\alpha^{\gamma_2n}$,  
the factor involving $(1-q(J))$ goes to one at large $n$.  Indeed,  $-q-q^2\le\ln(1-q)\le -q$ for $0<q<\frac{1}{2}$,
so  for $q$ of the form (\ref{q4.eq}),
\begin{equation}
 \ln(1-q)^{bn^2} \sim n^{J+2}\alpha^{\ff_2 n}
\label{limit1.eq}
\end{equation}
which goes to zero for large $n$.  Using the fact that 
$\binom{\ff_1\ff_3 n^2}{R_J} \lesssim 2^{\gamma_1\gamma_3n^2}$ for any choice of $R_J$, we thus have
\begin{equation}
  \mP(N_J=R_J) \sim 2^{\beta n^2 - \gamma_2nR_J|\log_2\alpha|}
\label{KRprob.eq}
\end{equation}
for large $n$, {where $\beta=\gamma_1\gamma_3$ is independent of $n$}.
Since there are $\sim 2^{\frac{n^2}{4} + o(n^2)}$ three-layer causal
sets, the number of {causal sets in $\Omega_{n,3,\Gamma, \vec
    p}$ }  with $N_J=R_J$ is thus  {at most}
\begin{equation}
\mathcal{N}(N_J=R_J) \sim 2^{\beta' n^2 - \gamma_2nR_J|\log_2\alpha|}, 
\label{num.eq}
\end{equation}
where $\beta'=\frac{1}{4}+\gamma_1\gamma_3$. 

{Since $N_0 \propto n^2$ for the dominant $K=3$ layer causal sets}, for the terms $N_{J>0}$ to matter in the BDG action (\ref{bdg.eq}) in the large $n$
limit, they, too, would have to be at least of order $n^2$.  But from
(\ref{num.eq}), {the number of three-layer sets with $N_{{J
      >0}}\sim \kappa n^2$ is suppressed by a factor of
  $2^{-\kappa\gamma_2|\log_2\alpha|n^3}$. For fixed linking fractions $p^{(12)}$ and
  $p^{(23)}$—and therefore fixed
    $\alpha$—this becomes negligible at large n.  As shown in \cite{ccs}, one can take this argument further
and show that $N_{{J>0}}$ grows as $n$, with a calculable proportionality factor, but we will not need that here.

It is useful to develop a more intuitive understanding of this suppression.  When we ask for the number of
$J$-element intervals in a causal set, we are counting the intervals that have $J$ elements \emph{and no more}.
For an interval ${(e,e')}$ in a large layered causal set, there are a very large number of  paths that could potentially connect 
$e$ and $e'$, and we must exclude most of them.  The factor $\alpha^{\ff_2 n}$ in (\ref{q4.eq}) comes from this
exclusion  and leads in turn to the exponential suppression in the number (\ref{num.eq}).  We will see below
that for causal sets with more layers the same kind of suppression occurs.

{There is one subtlety to this result, however. We have
  looked at the limit in which the linking fractions $p^{(12)}$  and
  $p^{(23)}$ remain fixed. In a layered set, though, the linking
  fraction can go  to zero as $n \rightarrow \infty$. If we allow  $p^{(12)}$  and
   $p^{(23)}$ to decrease fast enough as $n$  increases, the
   suppression of terms $N_{J>0}$ in Eqn.\eqref{num.eq} will no longer
  occur. In other words, as we increase the number of elements we can
  simultaneously make the links between adjacent layers sparser in a
  way that may allow  an
   appreciable number of small $J>0$ intervals}.
 Sets of this sort, however, are very rare. Consider two adjacent
 layers with a total of $\gamma n$ points. It follows from \cite{lc}
 that for $\delta \ll  1$, the number of possible arrangement of links
 between those layers with linking fraction less than $\delta$  is 
\begin{equation}
   \mathcal{N}(p < \delta) \sim 2^{\frac{\gamma^2 n^2}{4} (2\delta|\log_2 2\delta|)}
   \end{equation} 
while the number of possible arrangement of links with an arbitrary linking fraction
$\gamma^2n^2$  goes as $2^{\frac{\gamma^2 n^2}{4}}$ . As $n \rightarrow
\infty$, the sets with “sparse” layers thus become negligible, with
a measure approaching zero. Note in particular that the KR orders, as
originally defined in \cite{kr}, almost all have linking fractions
near $\frac{1}{2}$.

Combined with the results of \cite{ams}, this means that in the
$n\rightarrow \infty$ limit the BDG path integral suppresses all of the three-layer causal sets except
possibly a set of
measure zero.   We will discuss this remaining set of measure zero in
the conclusion. 
}

\section{Generalisation \label{main.sec}}

\begin{figure}[!bhtp]
\begin{center}
\begin{tikzpicture}[style=thick]
  \draw[line width=1pt, dashed] (-2,0) -- (5,0);
  \draw[line width=1pt, dashed] (-2,1) -- (5,1);
  \draw[line width=1pt, dashed] (-2,2) -- (5,2);
  \draw[line width=1pt, dashed] (-2,3) -- (5,3);
 
  \draw[] (3,0) -- (2,1);
  \draw[] (3,0) -- (4,1);
  \draw[] (2,1) -- (3,2);
\draw[] (4,1) -- (3,2);  
\draw[] (3,2)--(3,3); 

  \draw[] (0,0) -- (1,1);
  \draw[] (0,0) -- (-1,1);
  \draw[] (1,1) -- (1,2);
\draw[] (-1,1) -- (-1,2);  
\draw[] (-1,2)--(0,3);
\draw[] (1,2)--(0,3); 

\draw[] (3,0) node[label=below:$e_c$] {};
\draw[] (2,1) node {};
\draw[] (4,1) node {};
\draw[] (3,2) node {};
\draw[] (3,3) node [label=above:$e_d$] {};
\draw[] (0,0) node [label=below:$e_a$] {};
\draw[] (1,1) node {};
\draw[] (-1,1) node {};
\draw[] (-1,2) node {};
\draw[] (1,2) node {};
\draw[] (0,3) node [label=above:$e_b$] {};

\end{tikzpicture}
\caption{Two causal set intervals ${(e_a, e_b)}$ and ${(e_c,e_d)}$   with $(\ccL^3)_{ab}=(\ccL^3)_{cd}=2$, but
  $|{(e_a,e_b)}|=4$ while $|{(e_c,e_d)}|=3$.}\label{L32.fig} 
\end{center}
\end{figure}
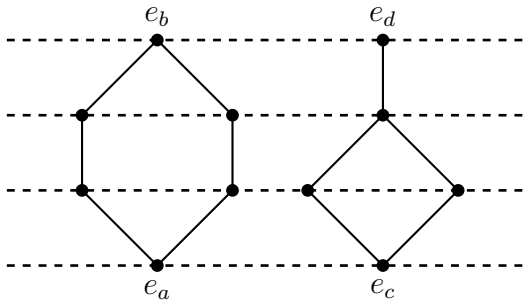

For general $K$, the calculation becomes more complicated. In particular, the
information contained in $({\ccL^r})_{ij}$ is not sufficient to infer the size of the interval $(\eki,
\ekpj)$. Fig.\ \ref{L32.fig} shows an example for $K=4$, in which the paths shown are all $2$-element paths obtained
from $\ccL^3$.  It is evident from the figure that $(\ccL^3)_{ab}=(\ccL^3)_{cd}=2$---that is, there are exactly two 
$4$-element paths starting from $e_a$ and ending at $e_b$, and also exactly  two $4$-element paths starting from $e_c$ and
ending at $e_d$ ---but
the interval sizes differ, $|{(e_a,e_b)}|=4$ and $|{(e_c,e_d)}|=3$.  Thus, one has to be  careful in extracting
$N_J$ from the $\ccL^r$ for $K>3$.  
 
\subsection{Strategy \label{strategy.ssec}}  

The BDG action $\SBDG(C)$ of (\ref{bdg.eq}) is a function of the $N_J$, the number of intervals in $C$ of size $J$.
The basic question we wish to ask is whether, for a $K$-layer causal set, the $N_{J>0}$ contribute significantly 
to the action or whether, as in the three-layer case of section \ref{secK3.ssec}, they become negligible 
for large $n$.  As in the three-layer case, we will first calculate the probability $\cP(N_J=R_J)$ of having exactly 
$R_J$ $J$-element intervals.  We will again find that the number of $K$-layer causal sets for which 
$R_J\sim\kappa n^2$---that is, the number that contribute significantly to the BDG action---is
again very highly suppressed for large $n$.   {It is worth emphasizing that this result is particular to {$K$-layered}
sets with $K\ll n${;} for sets obtained from a Poisson sprinkling of  $d$-dimensional Minkowski spacetime, the numbers $N_J$ all
grow as {$n^{2-\frac{2}{d}}$} for large $n$ \cite{glasersurya}}.

\begin{figure}[!bhtp]
\begin{center}
\begin{tikzpicture}[style=thick, pin distance=5.5ex]
  \draw[line width=1pt, dashed] (-6.5,0) -- (6,0);
  \draw[line width=1pt, dashed] (-6.5,1) -- (6,1);
  \draw[line width=1pt, dashed] (-6.5,2) -- (6,2);
  \draw[line width=1pt, dashed] (-6.5,3) -- (6,3);
  \draw[] (-5,0) -- (-6,1);
  \draw[] (-5,0) -- (-5,1);
  \draw[] (-5,0) -- (-4,1);
  \draw[] (-5,2) -- (-6,1);
  \draw[] (-5,2) -- (-5,1);
  \draw[] (-5,2) -- (-4,1);

 \draw[] (-2,0) -- (-3,1);
  \draw[] (-2,0) -- (-1,1);
  \draw[] (-3,1) -- (-2,3);
  \draw[] (-1,1) -- (-1,2);
  \draw[] (-1,2) -- (-2,3);

  \draw[] (1,0) -- (0,1); 
  \draw[] (1,0) --(2,1);
   \draw[] (0,1) -- (1,2); 
 \draw[] (2,1) -- (1,2);
 \draw[] (1,2) -- (1,3);

  \draw[] (4,0) -- (4,1); 
  \draw[] (4,1) --(3,2);
   \draw[] (4,1) -- (5,2); 
 \draw[] (3,2) -- (4,3);
 \draw[] (5,2) -- (4,3);

\draw[] (-5,0) node[label=below:$(a)$] {};
\draw[] (-6,1) node {};
\draw[] (-5,1) node {};
  \draw[] (-4,1) node {};
 \draw[] (-5,2) node {};
 
 \draw[] (-2,0) node[label=below:$(b)$] {};
 \draw[] (-3,1) node {};
 \draw[] (-1,1) node {};
  \draw[] (-1,2) node {};
 \draw[] (-2,3) node {};
   
\draw[] (1,0) node[label=below:$(c)$] {};
 \draw[] (0,1) node {};
  \draw[] (2,1) node {};
  \draw[] (1,2) node {};
  \draw[] (1,3) node {};
 
 \draw[] (4,0) node[label=below:$(d)$]  {};
 \draw[] (4,1) node {};
 \draw[] (3,2) node {};
 \draw[] (5,2) node {};
 \draw[] (4,3) node {};

 \end{tikzpicture}
\caption{Examples of different types of $3$-element intervals  in a
  $(K=4)$-layer causal set. While there are five possible unlabelled
  $3$-element causal sets, only four can be realised as 
  intervals for $K=4$. The $3$-element chain
  cannot be realised since it needs $K\geq 5$.} \label{figdiffcs.fig} 
\end{center}
\end{figure}
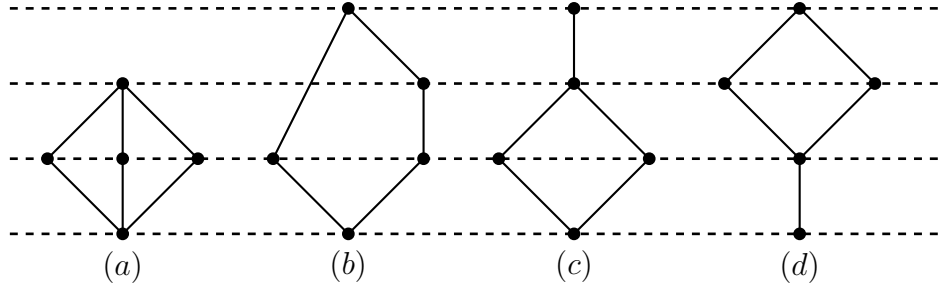

  To demonstrate our claimed suppression, consider a $J$-element interval  in a causal set $C$.  
  Each such interval is itself a
  $J$-element causal set $c \in \Omega_J$. {In the $K=3$ case above the $J$-element antichain is the only
    possible 
     interval, but for $K>3$ many other possibilities exist.}  As in Fig.\  \ref{ints.fig}(c), different
  $c \in \Omega_J$ can contribute to $N_J$, and this ``fine structure'' provides an important characterisation of
  $C$. {An example of such fine structure is given in Fig.\ \ref{figdiffcs.fig}}.

 In what follows, we will show that for large $n$, the contribution to
 $N_J$ for a typical $K$-layer causal set is
 dominated by a particular subset of $\Omega_J$, the $J$-antichains whose elements occupy a single layer.
While our main interest is in small values of $J$---for instance, $J=1,2,3$ for the BDG action in  $d=4$---we will 
show a much more general result as long as $J$ and $K$ remain fixed.  Our basic strategy is to split the calculation 
of the probability  $\cP(N_J=R)$ into three stages, starting with the types of causal sets that can occur as 
 intervals and successively refining their description.  Each stage, though, will be based on the same mathematical
structure, which we now summarize.

Suppose we have a set of objects that can come in $M$ ``colors'' indexed by $\alpha=1,\dots,M$.\footnote{For
    example, take the set of objects to be the set of all  $J$-element intervals, each ``colored'' by  which $J$-element causal
set in $\Omega_J$ it is.}  In  order to
pick some number $R$ of objects, we must have $R^{(1)}$ objects of color $1$, $R^{(2)}$ of color $2$, etc., with
\begin{equation}
\sum_{\alpha=1}^M \RJa = R, \qquad \RJa\ge0  .
\label{composition0}
\end{equation}
A sum of this form is called a \emph{weak composition}  of $R$ into $M$ parts, and is an example of what is sometimes
known in combinatorics as a walls and balls problem (``divide a row of $R$ balls into segments using $M-1$
walls'').  Note that some of the $\RJa$ can be zero.  Denote the set of such weak compositions of $R$
as $\Pi(R)$, where the number of parts is implicit in the notation.

Now suppose instead that the numbers of objects of color $\alpha$ is a random variable $\NJa$,
for which the value $\RJa$ occurs with probability $\cP(\NJa=\RJa)$.  It is then clear that
\begin{equation}
\cP(N=R) =\sum_{\{\RJa\}\in\Pi(R)}\left( \prod_{\alpha=1}^M \cP(\NJa=\RJa)\right), 
\label{NJK0.eq} 
\end{equation} 
{where $\{\RJa\}$ denotes a particular weak composition in the set of weak compositions $\Pi(R)$ of $R$.}  This
will be the fundamental equation we use for each stage.

In general, this expression for $\cP(N=R)$ is quite hard to calculate.  For the setting we are
interested in, however, it is possible to find a useful bound.  Note first that the number of 
weak compositions $\Pi(R)$ is $\binom{R+M-1}{R}$, which reaches its maximum at $R=M-1$, with
\begin{equation}
\binom{R+M-1}{R} \lesssim \frac{2^{2R}}{\sqrt{\pi R}}  .
\end{equation}
Hence 
\begin{equation}
\cP(N=R) \lesssim \frac{2^{2R}}{\sqrt{\pi R}}\, 
   \max_{\{\RJa \}\in\Pi(R)}\left( \prod_{\alpha=1}^M \cP(\NJa=\RJa)\right)  .
\label{Pmax.eq}
\end{equation} 
We will see below that the probabilities relevant to counting intervals in a causal set
take the general form
\begin{equation}
\cP(\NJa=\RJa) \sim 2^{-\beta_\alpha n^{r_\alpha}\RJa}  ,
\label{asymp.eq}
\end{equation}
where $\beta_\alpha$ and $r_\alpha$ are nonnegative constants and $n$ is the cardinality of the
set, which we will take to be very large.  Without loss of generality, we can choose the 
indexing to order the exponents $r_1 > r_2>r_3>\dots$ (the case of two or more exponents being equal  
is a trivial generalization).  Then for weak compositions for which $R^{(1)}\ne0$, the $\alpha=1$ 
term will dominate the product in (\ref{NJK0.eq}), with
$$\prod_{\alpha=1}^M \cP(\NJa=\RJa) \sim 2^{-\beta_1 n^{r_1}R^{(1)}}$$
For weak compositions in which $R^{(1)}=0$, on the other hand, the $\alpha=2$ term will
dominate, with
$$\prod_{\alpha=2}^M \cP(\NJa=\RJa) \sim 2^{-\beta_2
  n^{r_2}R^{(2)}}.$$
 Continuing this process, we see that the maximum in (\ref{Pmax.eq}) will be attained when 
$R^{(1)}=R^{(2)}=\dots=R^{(M-1)}=0$ and $R^{(M)}=R$, with
\begin{equation}
\max_{\{\RJa\}\in\Pi(R)}\left( \prod_{\alpha=1}^M \cP(\NJa=\RJa)\right)
    \sim 2^{-\beta_M n^{r_M}R}
\label{Pmax2.eq}
\end{equation}
where $r_M$ is the smallest of the exponents in the probabilities (\ref{asymp.eq}).   The problem 
thus reduces to a computation of these probabilities.

We can now describe the three stages of our calculation:

{\bf Stage 1:}\ Let $C$ be a randomly chosen causal set from $\OnKGp$.  We are interested in 
the number of $J$-element intervals in $C$, or more precisely in the probability $\cP(N_J=R)$ that $C$ contains 
precisely $R$ such intervals.   As noted above, 
each such interval is itself a $J$-element causal 
set.  List the members $\cJa\in\Omega_J$  {that are ``allowed''  
  $J$-element intervals in some
arbitrary but fixed order, $(c^{(1)},c^{(2)},\dots,c^{(M)})$, where $M\leq |\Omega_J|$.  For example, for $K=3$ the 
only 
possible 
 $J$-element interval is an antichain $\achain_J$, so for this case $M=1$  and $c^{(1)}=\achain_J$.  We
therefore need to consider all weak compositions of $R$ into $M$ parts.} 
Then, by the argument above, the probability that $C$ contains exactly $R$ $J$-element intervals is
\begin{equation}
\cP(N_J=R) =\sum_{\{\RJa\}\in\Pi(R)}\left( \prod_{\alpha=1}^{{M}}
  \cP(\NJa=\RJa)\right), 
\label{NJK.eq} 
\end{equation} 
 {where $\RJa$ now denotes the number of times $\cJa$ appears as 
 an interval in $C$.}  

To compute the probability (\ref{NJK.eq}),we need the set $\Omega_J$ of $J$-element causal sets, the set of weak
compositions of $R$, and the individual probabilities $\cP(\NJa=\RJa)$.  For the first, causal sets are completely
classified up to $J=16$ \cite{Brinkmann}; above that their number
grows very rapidly. However, we will see below that the
counting is dominated by a small subset of all causal sets.  For the second, the weak compositions of an integer are
well understood \cite{Wagner}.  We therefore focus on the third problem, the probabilities $\cP(\NJa=\RJa)$.

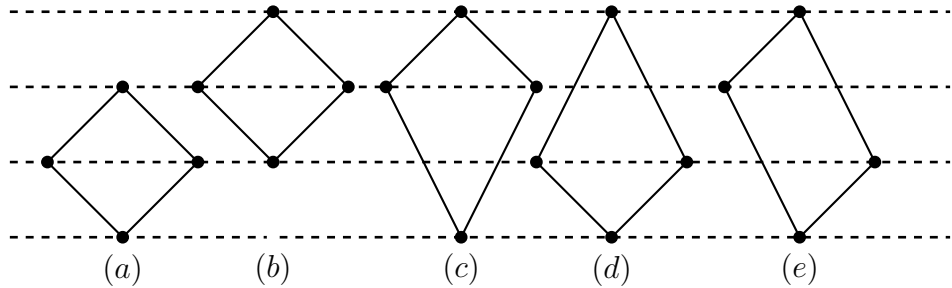
\begin{figure}[!bhtp]
\begin{center}
\begin{tikzpicture}[style=thick, pin distance=5.5ex]
  \draw[line width=1pt, dashed] (-6.5,0) -- (6,0);
  \draw[line width=1pt, dashed] (-6.5,1) -- (6,1);
  \draw[line width=1pt, dashed] (-6.5,2) -- (6,2);
  \draw[line width=1pt, dashed] (-6.5,3) -- (6,3);
  \draw[] (-5,0) -- (-6,1);
  \draw[] (-5,0) -- (-4,1);
  \draw[] (-5,2) -- (-6,1);
  \draw[] (-5,2) -- (-4,1);

 \draw[] (-3,1) -- (-4,2);
  \draw[] (-3,1) -- (-2,2);
  \draw[] (-3,3) -- (-4,2);
  \draw[] (-3,3) -- (-2,2);

  \draw[] (-.5,0) -- (-1.5,2);
  \draw[] (-.5,0) -- (.5,2);
  \draw[] (-.5,3) -- (-1.5,2);
  \draw[] (-.5,3) -- (.5,2);

  \draw[] (1.5,0) -- (.5,1);
  \draw[] (1.5,0) -- (2.5,1);
  \draw[] (1.5,3) -- (.5,1);
  \draw[] (1.5,3) -- (2.5,1);

  \draw[] (4,0) -- (3,2);
  \draw[] (4,0) -- (5,1);
  \draw[] (4,3) -- (3,2);
  \draw[] (4,3) -- (5,1);


\draw[] (-5,0) node[label=below:$(a)$] {};
 \draw[] (-6,1) node {};
  \draw[] (-4,1) node {};
 \draw[] (-5,2) node {};

 \draw (-3,0) node[fill=white,draw=white, label=below:$(b)$]{};
 \draw[] (-3,1) node {};
 \draw[] (-4,2) node {};
 \draw[] (-2,2) node {};
 \draw[] (-3,3) node {};
   
\draw[] (-.5,0) node[label=below:$(c)$] {};
 \draw[] (-1.5,2) node {};
  \draw[] (.5,2) node {};
 \draw[] (-.5,3) node {};

 \draw[] (1.5,0) node[label=below:$(d)$]  {};
 \draw[] (.5,1) node {};
 \draw[] (2.5,1) node {};
 \draw[] (1.5,3) node {};

\draw[] (4,0) node[label=below:$(e)$] {};
 \draw[] (3,2) node {};
  \draw[] (5,1) node {};
 \draw[] (4,3) node {};
 
 \end{tikzpicture}
\caption{{Five  embeddings of a two-element antichain as 
    an interval in a four-layer
causal set. }} \label{figx.fig} 
\end{center}
\end{figure}

{\bf Stage 2:}\ As the next stage, consider a particular {allowed}  $J$-element causal set $\cJa \in \Omega_J$.  This 
set can occur as 
 an interval in many different ways;
Fig.\ \ref{figx.fig}, for instance, shows five  
different embeddings of the two-element antichain as 
 an interval in a four-layer causal set.

As a first refinement, note that an interval  
 $\cJa$ may begin at any layer $k$
and end at any later layer $k'$, as long as $k'-k \ge h(c^\alpha)+1$, where $h(c^{\alpha})$ is the height of
$\cJa$ (as defined in section \ref{prelim.sec}).  In other words,  $\RJa$  receives
contributions from pairs of layers $(k, k')$ such that $(\eki,\ekpj)=\cJa, \, \eki \in \mL_k, \, \ekpj \in \mL_{k'}$.
If we denote the number of such contributions as $\RJakkp$, we again have a weak composition:
\begin{equation}
 \sum_{(k,k')\in\kappa} \RJakkp = \RJa, \qquad \RJakkp\ge0
\label{composition1}
\end{equation}
where $\kappa$ is the set of pairs $(k,k')$ with $k\in\{1,\dots,K-h(c^{\alpha})-1\}$ and $k'\in\{k+h_\alpha+1,\dots,K\}$.
Then, just as in stage 1, 
\begin{equation}
\cP(\NJa=\RJa) =\sum_{\{\RJakkp\}\in\Pi(\RJa)}
     \left( \prod_{(k,k')\in\kappa} \cP\left(\NJakkp=\RJakkp\right)\right).
\label{NJK2.eq} 
\end{equation} 
 
{\bf Stage 3:}\ Even after we have sorted the sets $\cJa$ by their initial and final layers, there is
further fine structure.  In Fig.\ \ref{figx.fig}, for instance, the intervals $(c)$, $(d)$, $(e)$, and $(f)$ all
start at layer $1$ and end at layer $4$, contributing to ${R^{(\alpha,1,4)}}$.   For a further refinement, we now
fix $k$ and $k'$ and consider embeddings $\cJakkp \hookrightarrow
C$, where $\cJakkp $ is a realisation of $ \cJa$ for an
  interval between $\mL_k$ and $\mL_{k'}$.  Each
embedding will determine a set of numbers $\vec n = (n_{k_1}, \ldots n_{k_r})$, where $n_{k_a}$ is 
the  the number of elements of $\cJakkp$ lying in the layer $k_a$, with the 
ordering  $k < k_1  < \ldots < k_r <k'$.   The intervals with each $\vec n$, in turn,  will contribute a 
number $\RJakkpn$ to $\RJakkp$.  We thus have another weak composition,
with
\begin{equation}
 \sum_{\vec n_c \in \mathcal E_\alpha^{(k,k')} }\RJakkpn = \RJa, 
      \qquad \RJakkpn \ge 0
\label{composition2}
\end{equation}
where $\mathcal E_\alpha^{(k,k')}$ is the set of possible embeddings $\vec n$ of $\cJa$ in the 
intervening $k'-k-1$ layers, and
\begin{equation}
\cP(\NJakkp=\RJakkp) = \sum_{\{\RJakkpn\}\in\Pi(\RJakkp)}
 \left(\, \prod_{\vec n_c \in \mathcal E_\alpha^{(k,k')}}\cP\left(\NJakkpn=\RJakkpn\right)\right).
\label{NJK3.eq} 
\end{equation} 
 
Now let $q(\vec n_\alpha)$ 
denote the probability of a particular set $\cJa$ occurring as 
an interval between layers 
$k$ and $k'$ with an internal pattern given by $\vec c$---that is, the probability of an interval
$(\eki,\ekpj)=\cJa, \eki \in \mL_k, \ekpj \in \mL_{k'}$ with embedding $\vec n_\alpha$.  Then as we saw 
in the $K=3$ case, 
\begin{equation}
\cP(N_J^{(\alpha,k,k',\vec n_\alpha)}=R) = 
  \binom{\gamma_k \gamma_{k'} n^2}{R} q(\vec n_\alpha)^R (1-q(\vec n_\alpha))^{\gamma_k
  \gamma_{k'} n^2 -R}.
\label{Pvecn.eq}
\end{equation}

To evaluate  this expression, we face the daunting task of calculating $q(\vec n_\alpha)$ for each 
$\vec n_\alpha$ associated with each pair of layers  $(k,k')$ and each $\cJa \in \Omega_J$.  To do so 
explicitly for general $J$ and $K$ is highly nontrivial: the number of
possible intervals 
 in stage 1 grows as
$2^{J^2/4}$, for example.  

Note, however, that if a particular set of probabilities dominates in the large $n$ limit, we can ignore the others 
to leading order in sums over weak compositions.  In particular, for
probabilities of the form (\ref{asymp.eq})  the sum over compositions is bounded by the much simpler expression  (\ref{Pmax2.eq}).
 
A further simplification comes from the structure of the pairs $(k,k')$ in Stage 2.  Suppose we have an interval
$(\eki,\ekpj)$ whose initial element lies in a layer $k>1$.  By definition, the interval depends only on elements to the
future of $e$, and thus to the future of layer $k$.  This means that in calculating probabilities, we can ignore
all the layers that come before $\mL_k$.  Similarly, we can ignore the layers that come after $\mL_{k'}$.  
This reduces the calculation to one for a causal set with fewer layers.

\subsection{Four layers ($K=4$) \label{mainK4.ssec}}    

We now explicitly work out the four-layer case for $J=1,2,3$. This provides the necessary intuition for 
generalisation to arbitrary $K$ and $J$.  The reader may at any point skip to the general case,
section \ref{mainKJ.ssec}.

\subsubsection{$J=1$ \label{J1.sec}}

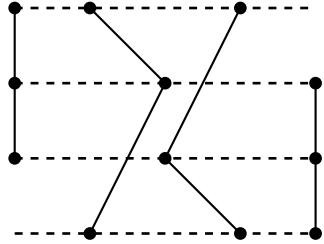
\begin{figure}[!bhtp]
\begin{center}
\begin{tikzpicture}[style=thick]
  \draw[line width=1pt, dashed] (0,0) -- (4,0);
  \draw[line width=1pt, dashed] (0,1) -- (4,1);
  \draw[line width=1pt, dashed] (0,2) -- (4,2);
  \draw[line width=1pt, dashed] (0,3) -- (4,3);
\draw[color=black] (4,0) -- (4,1);
\draw[color=black] (4,1) -- (4,2);
\draw[color=black] (3,0) -- (2,1);
\draw[color=black] (2,1) -- (3,3);
\draw[color=black] (1,0) -- (2,2);
\draw[color=black] (2,2) -- (1,3);
\draw[color=black] (0,1) -- (0,2);
\draw[color=black] (0,2) -- (0,3);
\draw[color=black] (4,0) node {};
\draw[color=black] (4,1) node {};
\draw[color=black] (4,2) node {};
\draw[color=black] (1,0) node {};
\draw[color=black] (2,2) node {};
\draw[color=black] (1,3) node {};
\draw[color=black] (3,0) node {};
\draw[color=black] (2,1) node {};;
\draw[color=black] (3,3) node {};
\draw[color=black] (0,1) node {};
\draw[color=black] (0,2) node {};
\draw[color=black] (0,3) node {};
\end{tikzpicture}
\caption{{For $K=4, J=1$, there are different choices of pairs $\{(k,k')\}$, and therefore also different choices of
    embeddings. For the pair $(1,3)$, there is only one embedding, in which $e$ lies in layer $2$.  For the
    pair $(2,4)$  there is again only one choice, in which $e$ lies in layer $3$. For the
    pair $(1,4)$, on the other hand, $e$ can lie in either layer $2$ or layer $3$. All four possibilities must 
    be considered when calculating $\cP(N_1=R)$.}}\label{Kfourembed.fig} 
\end{center}
\end{figure}

Fig.\  \ref{Kfourembed.fig} illustrates the various possible types of $J=1$ intervals when $K=4$. $\Omega_{J=1}$ consists of the single
element $e$, hence for Stage 1 the $|\Omega_1|$  partition of $R$ is trivial (Eqn.\ \eqref{NJK.eq}). Next, for Stage 2,  
$h(c)=1$, hence we require $k' \geq k+2$.  This gives us the  possible pairs $\kappa = \{(1,3),(2,4),(1,4) \} $.  
For Stage 3, the pairs $(1,3) $ and $(2,4)$ each have only one possible embedding each:  $\vec n= (n_2=1)$ 
and $\vec n=(n_3=1)$, respectively.  The pair $(1, 4)$, on the other hand, has two possible embeddings,
 $\vec n =(n_2=1) $ and $\vec n = (n_3=1)$.       

For $(k,k')=(1,3)$ or  $(2,4)$,  the analysis is the same as the $K=3$ case of section \ref{secK3.ssec}.  As
in Eqn.\ (\ref{q3.eq}), specialized to $J=1$, we have
\begin{equation}
q_{13} = \gamma_2 n \pi_{13} (1-\pi_{13})^{\gamma_2n -1}, \quad q_{24} = \gamma_3 n \pi_{24} (1-\pi_{24})^{\gamma_3n-1}
  \end{equation} 
  where  $\pi_{13} = p^{(12)}p^{(23)}$ and $\pi_{24} = p^{(23)}p^{(34)}$. In the large $n$ limit these behave asymptotically as
  \begin{equation}
    q_{13} \sim 2^{-\gamma_2n|\log_2\alpha_{13}|} ,
    \quad q_{24}  \sim 2^{-\gamma_3n|\log_2\alpha_{24}|}
  \end{equation}
 where, as in section \ref{secK3.ssec}, $\alpha_{ij} = 1-\pi_{ij}$.

 For the pair $(1,4)$, we can see from Fig.\ \ref{Kfourembed.fig} that there are two embeddings, which we can
 label $\vec n_2=(n_2=1,n_3=0)$ and $\vec n_3 =  (n_2=0,n_3=1)$.  Let us first consider $\vec n_2$.  To obtain
 a probability for this configuration, for a fixed initial element $e_i^{(1)}\in\mL_1$ and a fixed final element $e_j^{(4)}\in\mL_4$, we 
 must require that
 \begin{itemize}
 \item There is exactly one $3$-element path from $e_i^{(1)}$ to $e_j^{(4)}$ via  an element in $\mL_2$. This contributes a factor 
 $\gamma_2 n \pi_{124} (1-\pi_{124})^{\gamma_2n -1}$, where $\pi_{124}=  p^{(12)}p^{(24)}$.
 \item There is \emph{no}  path from $e_i^{(1)}$ to $e_j^{(4)}$ via  an element  in $\mL_3$,  since ${(e_i^{(1)}, e_j^{(4)})}$ would then contain more than
 one element.  This contributes a factor  $(1-\pi_{134})^{\gamma_3n}$, where $\pi_{134}=  p^{(13)}p^{(34)}$.
 \item There is \emph{no}  $4$-element path from $e_i^{(1)}$ to $e_j^{(4)}$ via an element in $\mL_2$ and an element in $\mL_3$, since again
 ${(e_i^{(1)}, e_j^{(4)})}$ would contain more than one element.   Note that this condition is equivalent to a 
 restriction on the link matrix, $(\ccL^3)_{ij}=0$.  Since there are $\gamma_2\gamma_3n^2-\gamma_3n$  pairs of elements 
 $(e_{i'}^{(2)}\in \mL_2, e_{j'}^{(3)} \in \mL_3)$ not including the
 element in the $3$-element path,  this contributes a factor of $(1-\pi_{1234})^{\gamma_2\gamma_3n^2-\gamma_3n}$, 
 where $\pi_{1234}= p^{(12)}p^{(23)}p^{(34)}$.  
 \end{itemize}
 The calculation for the second embedding, $\vec n_3$, is identical, except that layers $2$ and $3$ are switched.  
 Thus
 \begin{eqnarray} 
q_{124} &=&  \gamma_2 n \pi_{124} 
   (1-\pi_{124})^{\gamma_2n -1} (1-\pi_{134})^{\gamma_3 n} (1-
            \pi_{1234})^{\gamma_2\gamma_3n^2-\gamma_3n} \nonumber \\
q_{134} &=&  \gamma_3 n \pi_{134}
(1-\pi_{134})^{\gamma_3n -1}(1-\pi_{124})^{\gamma_2n} (1-\pi_{1234})^{\gamma_2\gamma_3n^2-\gamma_2n} , 
    \end{eqnarray} 
 The last term in each expression clearly dominates in the large $n$ limit, giving identical leading asymptotic behavior,
 \begin{equation} 
q_{124} \sim q_{134} \sim  2^{-\gamma_2\gamma_3n^2|\log_2\alpha_{1234}|} 
 \end{equation} 
 with $\alpha_{1234} = 1 - \pi_{1234}<1$.

This example illustrates an  important feature: the suppression is enhanced when the number of intervening layers
increases.   More layers allow more paths between the endpoints of an interval, which must be excluded to keep
the interval size fixed.  In fact, as we have seen, more layers allow \emph{many} more paths, since one can independently
specify elements in each layer.

Applying Eqn.\ \eqref{Pvecn.eq} to the two partitions $\vec n_2$ and $\vec n_3$ of $(1,4)$, we have 
\begin{eqnarray}
 \cP(N_1^{(1,4,\vec n_2)}=R) &=  \binom{\gamma_1 \gamma_{4} n^2}{R} q_{124}^R (1-q_{124})^{\gamma_1  \gamma_4 n^2 -R}  
                               \sim   2^{\beta_{(1,4)} n^2 - \gamma_2\gamma_3n^2R|\log_2\alpha_{1234}|}
                                \nonumber \\
  \cP(N_1^{(1,4,\vec n_3)}=R) &= \binom{\gamma_1 \gamma_{4} n^2}{R} q_{134}^R (1-q_{134})^{\gamma_1  \gamma_4 n^2 -R}  
                               \sim  2^{\beta_{(1,4)} n^2 - \gamma_2\gamma_3n^2R|\log_2\alpha_{1234}|}
\end{eqnarray} 
where $\beta_{(1,4)}=\gamma_1\gamma_4$. Here we have used the facts that 
$\binom{\gamma_1 \gamma_{4} n^2}{R} \lesssim 2^{\gamma_1 \gamma_{4} n^2}$ and that both
$(1-q_{124})^{\gamma_1  \gamma_4 n^2}$ and $(1-q_{134})^{\gamma_1  \gamma_4 n^2}$ go to $1$ for large $n$, as 
in Eqn.\ (\ref{limit1.eq}).  Inserting into  Eqn.\ \eqref{NJK3.eq},
\begin{eqnarray} 
  \cP(N_1^{(1,4)}\!=R)  &=& \!\!\!\!  \sum_{R^{(1,4,\vec n_2)}+R^{(1,4,\vec n_3)}=R  } \!\! \cP(N_1^{(1,4,\vec
  n_2)}\!=R^{(1,4,\vec n_2)}) \cP(N_1^{(1,4,\vec n_3)}\!=R^{(1,4,\vec n_3)}) \nonumber \\ 
   &\sim&  2^{\beta_{(14)}n^2- \gamma_2\gamma_3n^2R|\log_2\alpha_{1234}|} . 
  \label{14.eq} 
\end{eqnarray} 
Since both $(1,3)$ and $(2,4)$ admit just a single partition,  from Eqn.\ \eqref{Pvecn.eq}  
\begin{eqnarray}
\cP(N_1^{(1,3)}=R) &=  \binom{\gamma_1 \gamma_{3} n^2}{R} q_{13}^R (1-q_{13})^{\gamma_1  \gamma_3 n^2 -R}   
                               \sim  2^{\beta_{(13)}n^2- \gamma_2nR|\log_2\alpha_{13}|}           \nonumber \\
 \cP(N_1^{(2,4)}=R) &= \binom{\gamma_2 \gamma_{4} n^2}{R} q_{24}^R (1-q_{24})^{\gamma_2  \gamma_4 n^2 -R}   
                                \sim 2^{\beta_{(24)}n^2- \gamma_3 nR|\log_2\alpha_{24}|} , 
  \label{1324.eq} 
\end{eqnarray}
where $\beta_{(1,3)}=\gamma_1\gamma_3$ and $\beta_{(2,4)}=\gamma_2\gamma_4$.   
Comparing Eqns.\ \eqref{14.eq} and \eqref{1324.eq}, {we note that the leading  negative term
  in the exponent in the latter is of lower order in $n$ than the
  former,  and hence when $R \propto n^2$ (to
    leading order)  it dominates for any finite composition. This is a crucial feature of our calculation: when there are
  more layers in between $k$ and $k'$, i.e., $k'-k> h(c)+1$, the leading negative term in the exponent  is of higher order in $n$, and hence is suppressed
  relative to the ``tightest  fit'' $k'-k=h(c)+1$}.
Hence,  of the $|\kappa|$ compositions in Eqn.\ \eqref{composition1}, the dominant ones are those for which 
$R_{14} \sim 0$.  If ${\tilde\gamma} |\log_2 {\tilde\alpha}|$ is the smaller of  $\gamma_2 |\log_2\alpha_{13}|$  and 
$\gamma_3 |\log_2 \alpha_{24}|$, then 
\begin{equation}
\cP(N_1=R) \sim 2^{{\tilde\beta}n^2- {\tilde\gamma}nR |\log_2{\tilde\alpha}|}  .
 \end{equation} 
This is identical in form to Eqn.\ \eqref{KRprob.eq}: the analysis has thus reduced to that for  $K=3$.  

\subsubsection{$J=2$ \label{J2.sec}}

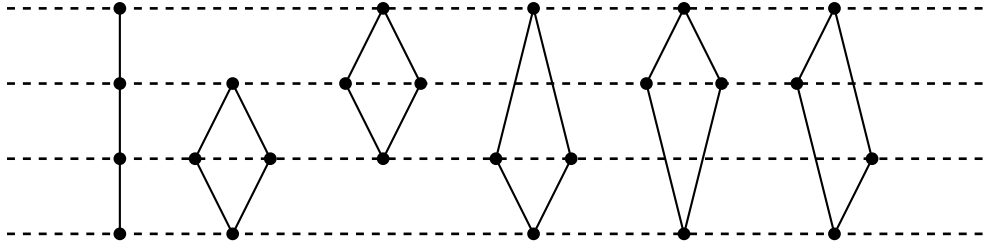
\begin{figure}[!bhtp]
\begin{center}
\begin{tikzpicture}[style=thick]
  \draw[line width=1pt, dashed] (-7,0) -- (6,0);
  \draw[line width=1pt, dashed] (-7,1) -- (6,1);
  \draw[line width=1pt, dashed] (-7,2) -- (6,2);
  \draw[line width=1pt, dashed] (-7,3) -- (6,3);
 
\draw[color=black] (-5.5,0) -- (-5.5,1);
\draw[color=black] (-5.5,1) -- (-5.5,2);
\draw[color=black] (-5.5,2) -- (-5.5,3);

\draw[color=black] (-4,0) -- (-4.5,1);
\draw[color=black] (-4,0) -- (-3.5,1);
\draw[color=black] (-4.5,1) -- (-4,2);
\draw[color=black] (-3.5,1) -- (-4,2);

\draw[color=black] (-2,1) -- (-2.5,2);
\draw[color=black] (-2,1) -- (-1.5,2);
\draw[color=black] (-2.5,2) -- (-2,3);
\draw[color=black] (-1.5,2) -- (-2,3);

\draw[color=black] (0,0) -- (-0.5,1);
\draw[color=black] (0,0) -- (0.5,1);
\draw[color=black] (-0.5,1) -- (0,3);
\draw[color=black] (0.5,1) -- (0,3);

\draw[color=black] (2,0) -- (2.5,2);
\draw[color=black] (2,0) -- (1.5,2);
\draw[color=black] (2.5,2) -- (2,3);
\draw[color=black] (1.5,2) -- (2,3);

\draw[color=black] (4,0) -- (4.5,1);
\draw[color=black] (4,0) -- (3.5,2);
\draw[color=black] (4.5,1) -- (4,3);
\draw[color=black] (3.5,2) -- (4,3);

\draw[color=black] (-5.5,0) node {};
\draw[color=black] (-5.5,1) node {};
\draw[color=black] (-5.5,2) node {};
\draw[color=black] (-5.5,3) node {};

\draw[color=black] (-4,0) node {};
\draw[color=black] (-4.5,1) node {};
\draw[color=black] (-3.5,1) node {};;
\draw[color=black] (-4,2) node {};

\draw[color=black] (-2,1) node {};
\draw[color=black] (-2.5,2) node {};
\draw[color=black] (-1.5,2) node {};;
\draw[color=black] (-2,3) node {};

\draw[color=black] (0,0) node {};
\draw[color=black] (-0.5,1) node {};
\draw[color=black] (0.5,1) node {};;
\draw[color=black] (0,3) node {};

\draw[color=black] (2,0) node {};
\draw[color=black] (2.5,2) node {};
\draw[color=black] (1.5,2) node {};
\draw[color=black] (2,3) node {};

\draw[color=black] (4,0) node {};
\draw[color=black] (4.5,1) node {};
\draw[color=black] (3.5,2) node {};
\draw[color=black] (4,3) node {};
\end{tikzpicture}
\caption{{The six different ways in which the  2-element interval can
    be realised in a $K=4$ layer causal set. To the
    extreme left is the chain $\chain_2$, which has only one realisation, while the antichain $\achain_2$ can be
    realised in five different ways. } }\label{K4J2.fig} 
\end{center}
\end{figure}

$\Omega_{J=2}$ consists of two causal sets: the two-element antichain $\achain_2$ (\begin{tikzpicture}[] \draw[color=black] (0,0) node {};
\draw[color=black] (0.4,0) node {};. 
\end{tikzpicture}) 
  and   the two-element chain
$\chain_2$  ($\,$ \begin{tikzpicture}[baseline=-10pt]{{\draw[color=black] (0,-0.4) node {};
\draw[color=black] (0,0) node {}; \draw[color=black] (0,-0.4) -- (0,0);}}
\end{tikzpicture}$\,$ ).  Since {$h(\chain_2)=2$},  
for   $K=4$ there is only one choice of initial and final layer, $(k,k')=(1,4)$. For $\achain_2$, on the
other hand, we have the three possibilities: $\kappa=\{(1,3),(2,4),(1,4) \}$. The first two of these are similar to
the $K=3$ case; each admits a single embedding, $\vec n_2=(n_2=2)$ and $\vec n_3=(n_3=2)$. For the 
choice $(1,4)$, on the other hand,  there are three possible embeddings, shown in Fig.\ \ref{K4J2.fig}: 
$\mathcal{E} = \{(n_2=2), (n_3=2),(n_2=1,n_3=1) \}$.
We calculate $q(\vec n_c)$ for each of these possibilities:

We first consider the two-element chain $\chain_2$.  For this set, there is only one possible embedding: 
$(k,k')=(1,4)$ with $\vec n =(n_2=1, n_3=1)$.    To obtain a probability for this configuration, for a fixed initial 
element $e_i^{(1)}\in\mL_1$ and a fixed final element $e_j^{(4)}\in\mL_4$, we must require that
 \begin{itemize}
 \item There is exactly one $4$-element path from $e_i^{(1)}$ to $e_j^{(4)}$, which goes from $e_i^{(1)}$ to an element in $\mL_2$  to an element in  $\mL_3$
   and then to $e_j^{(4)}$.  This contributes a factor $\gamma_2\gamma_3 n^2 \pi_{1234} (1-\pi_{1234})^{\gamma_2\gamma_3n^2 -1}$, where 
 $\pi_{1234}=  p^{(12)}p^{(23)}p^{(34)}$.
 \item There is no $3$-element path from $e_i^{(1)}$ to $e_j^{(4)}$.  This contributes a factor $(1-\pi_{124})^{\gamma_2n-1}$
and a factor $(1-\pi_{134})^{\gamma_3n-1}$  (to exclude paths with 
an element in either $\mL_2$ or $\mL_3$ that is also not in the
$4$-element path), where $\pi_{124} = p^{(12)}p^{(24)}$ and $\pi_{134}=p^{(13)}p^{(34)}$.
\end{itemize}
Hence
\begin{equation}
q_{1234}=(\gamma_2\gamma_3 n^2)\,  \pi_{1234} (1-\pi_{1234})^{\gamma_2\gamma_3n^2-1} 
  (1-\pi_{124})^{\gamma_2n-1}(1-\pi_{134})^{\gamma_3n-1}  \sim 2^{-\gamma_2\gamma_3n^2|\log_2\alpha_{1234}|} ,
\end{equation} 
where again $\alpha_{1234} = 1-\pi_{1234}$.
The insight we gain  from this example is that for $(\eki, \ekpj)=c$,  the leading order behavior goes 
as  $2^{-\beta n^r}$, where $r=k'-k-1$.  {Thus, the  smaller $r$ is,  the more dominant is the term.} 
  
For the two-element antichain $\achain_2$, the possible embeddings are shown in Figs.\ \ref{figx.fig} and
\ref{K4J2.fig}. There are three possible $(k,k')$ values: $\kappa =\{(1,3), (2,4), (1,4) \}$. 
The first two  reduce to the the $K=3$ case 
\begin{eqnarray} 
  q_{13} &=& \binom{\gamma_2n}{2} \pi_{123}^2(1-\pi_{123})^{\gamma_2 n -2} 
         \sim  2^{-\gamma_2n|\log_2\alpha_{123}|}  \nonumber \\
  q_{24} &=& \binom{\gamma_3n}{2} \pi_{234}^2(1-\pi_{234})^{\gamma_3 n -2} 
          \sim  2^{-\gamma_3n|\log_2\alpha_{234}|}   
\label{tighta2.eq} 
\end{eqnarray} 
where again $\pi_{123}=p^{(12)}p^{(23)}$, $\pi_{234}=p^{(23)}p^{(34)}$, and $\alpha_{ijk} = 1-\pi_{ijk}$.

For $(k,k')=(1,4)$, the possible embeddings $\mathcal E$ are $\vec n_1\equiv (n_2=2,n_3=0)$, $\vec n_2 \equiv
(n_2=0,n_3=2)$, and $\vec n_3 \equiv (n_2=1,n_3=1)$.
Denote the associated probabilities by $q_{14}^{\vec n_1}$, $q_{14}^{\vec n_2}$, and $q_{14} ^{\vec n_3}$.  Consider
first $q_{14}^{\vec n_1}$
with an initial element $e_i^{(1)}\in\mL_1$ and a final element $e_j^{(4)}\in\mL_4$.  As in the previous cases, we must require that
\begin{itemize}
 \item There are exactly two $3$-element paths from $e_i^{(1)}$ to $e_j^{(4)}$ via two distinct elements in $\mL_2$.  This contributes 
 a factor $\binom{\gamma_2}{2} (\pi_{124})^2 (1-\pi_{124})^{\gamma_2n -2}$. 
 \item There is no $3$-element path from  $e_i^{(1)}$ to $e_j^{(4)}$ that passes  through an element in $\mL_3$. This contributes a factor $(1-\pi_{134})^{\gamma_3n}$
(to exclude all paths with an element in $\mL_3$).
\item There is no $4$-element path from  $e_i^{(1)}$ to $e_j^{(4)}$ via elements in $\mL_2$ and $\mL_3$.  This is again
equivalent to the requirement that $(\ccL^3)_{ij}=0$, and it
contributes a factor of $(1-\pi_{1234})^{\gamma_2\gamma_3n^2-2\gamma_3
n}$
to the probability.
\end{itemize}
For $q_{14}^{\vec n_2}$, and $q_{14} ^{\vec n_3}$ the arguments are almost identical.  We thus obtain
\begin{eqnarray} 
q_{14}^{\vec n_1} &=& \binom{\gamma_2 n}{2} \pi_{124}^2
             (1-\pi_{124})^{\gamma_2n-2}(1-\pi_{134})^{\gamma_3n}(1-\pi_{1234})^{\gamma_2\gamma_3n^2-2\gamma_3n}  
            \sim  2^{-\gamma_2\gamma_3 n^2 |\log_2\alpha_{1234}|} \nonumber \\[.5ex]
q_{14}^{\vec n_2}&= & \binom{\gamma_3 n}{2} \pi_{134}^2
             (1-\pi_{134})^{\gamma_3n-2}(1-\pi_{124})^{\gamma_2n}(1-\pi_{1234})^{\gamma_2\gamma_3n^2-2\gamma_2n} 
            \sim 2^{-\gamma_2\gamma_3 n^2 |\log_2\alpha_{1234}|} \label{loosea2.eq}   \\[.5ex]
q_{14}^{\vec n_3}&= & \gamma_2 \gamma_3 n^2   \pi_{124} \pi_{134} (1-\pi_{124})^{\gamma_2n-1}
             (1-\pi_{134})^{\gamma_3n-1}(1-\pi_{1234})^{\gamma_2\gamma_3n^2-\gamma_2n  - \gamma_3 n}  
            \sim 2^{-\gamma_2\gamma_3 n^2 |\log_2\alpha_{1234}|} . \nonumber     
\end{eqnarray}
We see that the leading $n$ dependence is completely dictated by the factor $(1-\pi_{1234})^{\gamma_2\gamma_3n^2} $,
the term that excludes ``extra'' $4$-element paths.  Comparing Eqns.\ \eqref{tighta2.eq} and \eqref{loosea2.eq} we can
see that, as in the $J=1$  case,  the  ``tight  fit'' $k'=k+2$,  configurations dominate over the ``loose fit'' configurations $k'>k+2$. 

 Putting these together, we have for  the $\chain_2$ contribution 
 \begin{equation} 
\cP(N_2^{(\chain_2)}=R) = \binom{\gamma_1 \gamma_4 n^2}{R} q_{1234}^R (1-q_{1234})^{\gamma_1\gamma_4 n^2-R} 
\lesssim 2^{\beta_{14}n^2-\gamma_2 \gamma_3 R  n^2 |\log_2 (\alpha_{1234})| } ,
\end{equation} 
where $\beta_{14}=\gamma_1\gamma_4$ and  we have again used the bound $\binom{M}{N}\lesssim\binom{M}{M/2}\lesssim2^M$.

The $\achain_2$ contributions are 
\begin{eqnarray} 
\cP(N_2^{(\achain_2,1,3)}=R)&=&\binom{\gamma_1 \gamma_3 n^2}{R} q_{13}^R (1-q_{13})^{\gamma_1\gamma_3 n^2-R}       
         \lesssim  2^{\beta_{13} n^2 -\gamma_2 n R |\log_2\alpha_{123}| }  \nonumber\\[.5ex]
 \cP(N_2^{(\achain_2,2,4)}=R)&=&\binom{\gamma_2 \gamma_4 n^2}{R} q_{24}^R (1-q_{24})^{\gamma_2\gamma_4 n^2-R} 
          \lesssim  2^{\beta_{24} n^2 -\gamma_3 n R |\log_2\alpha_{234}| }    ,
\label{Ptightfit.eq}
\end{eqnarray} 
where $\beta_{13}=\gamma_1\gamma_3, \beta_{24}=\gamma_2\gamma_4$, and
\begin{eqnarray} 
&&\cP(N_2^{(\achain_2,1,4,\vec n_1)}=R) = \binom{\gamma_1 \gamma_4 n^2}{R} (q_{14}^{\vec n_1})^R (1-q_{14}^{\vec n_1})^{\gamma_1\gamma_4 n^2-R}   
\lesssim  2^{\beta_{14}n^2 -\gamma_2 \gamma_3 n^2 R |\log_2\alpha_{1234}| }  
    \nonumber\\
&&\cP(N_2^{(\achain_2,1,4,\vec n_2)}=R) = \binom{\gamma_1 \gamma_4n^2}{R} (q_{14}^{\vec n_2})^R (1-q_{14}^{\vec n_2})^{\gamma_1\gamma_4 n^2-R}  
\lesssim  2^{\beta_{14}n^2 -\gamma_2 \gamma_3 n^2 R |\log_2\alpha_{1234}| }  
     \  \\
&&\cP(N_2^{(\achain_2,1,4,\vec n_3)}=R) = \binom{\gamma_1 \gamma_4 n^2}{R} (q_{14}^{\vec n_3})^R (1-q_{14}^{\vec n_3})^{\gamma_1\gamma_4 n^2-R}  
\lesssim  2^{\beta_{14}n^2 -\gamma_2 \gamma_3 n^2 R |\log_2\alpha_{1234}| } . \nonumber
 \end{eqnarray} 
Comparing the contributions, it is clear that $\cP(N_2=R)$ is dominated by the partition with  $R$ coming from the
tight fit contributions of $\achain_2$, Eqn.\ \eqref{Ptightfit.eq}. Thus 
\begin{equation}
\cP(N_2=R) \lesssim 2^{{\tilde\beta} n^2 - {\tilde\gamma} n R|\log_2{\tilde\alpha}| }  
  \end{equation} 
where ${\tilde\gamma}|\log_2{\tilde\alpha}|$  is the smaller of $\gamma_2|\log_2\alpha_{123}|$ and $\gamma_3|\log_2\alpha_{234}|$.
We again see  that the dominant contribution comes from the three-layer-like tight embeddings of antichains, and the probabilities
have the same structure as those in the three-layer case.
 
\subsubsection{$J=3$\label{J3.sec}}

Here we extend the analysis to the significantly more involved case  $J=3$.  This is a special case of the more general analysis of section 
\ref{mainKJ.ssec}, but the
details may help with intuition.

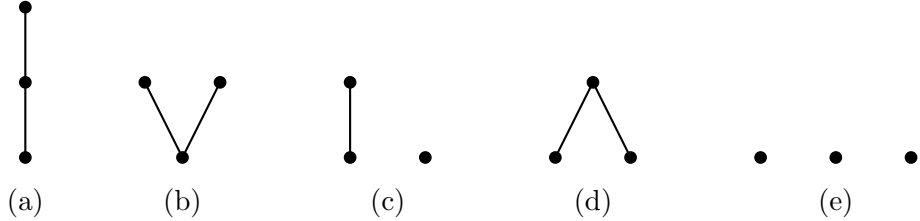
\begin{figure}[!bhtp]
 \begin{subfigure}[b]{.3\linewidth} 
\begin{center}
\begin{tikzpicture}[style=thick] 
  
  \draw (0,0) -- (0,1);
  \draw (0,1) -- (0,2);
  \draw (0,0) node[]  {} ;
  \draw (0,1) node {};
  \draw (0,2) node {}; \end{tikzpicture}
  \caption{}
  \end{center}
\end{subfigure}\hspace{-3cm} 
 \begin{subfigure}[b]{.3\linewidth} 
  \begin{center}
    \begin{tikzpicture}[style=thick]
      \draw (2,0) -- (1.5,1);
  \draw (2,0) -- (2.5,1);
 \draw (2,0) node   {};
  \draw (1.5,1) node {};
  \draw (2.5,1) node {};
  \end{tikzpicture}
  \caption{}
  \end{center}
\end{subfigure} \hspace{-2.5cm} 
\begin{subfigure}[b]{.3\linewidth} 
  \begin{center}
    \begin{tikzpicture}[style=thick]
      \draw (4,0) -- (4,1);
       \draw (4,0) node   {};
  \draw (4,1) node {};
  \draw (5,0) node {};
  \end{tikzpicture}
 \caption{}
  \end{center}
\end{subfigure} \hspace{-2.5cm} 
  \begin{subfigure}[b]{.3\linewidth} 
  \begin{center}
    \begin{tikzpicture}[style=thick]
\draw (6.5,0) -- (7,1);
\draw (7.5,0) -- (7,1);
\draw (6.5,0) node  {};
  \draw (7,1) node {};
  \draw (7.5,0) node {}; \end{tikzpicture}
   \caption{}
  \end{center}
\end{subfigure} \hspace{-2cm} 
 \begin{subfigure}[b]{.3\linewidth} 
  \begin{center}
    \begin{tikzpicture}[style=thick] 
 \draw (9,0) node {};
  \draw (10,0) node {};
  \draw (11,0) node {};
\end{tikzpicture}
 \caption{}
  \end{center}
\end{subfigure} 
\caption{{The set of unlabelled  $3$-element causal sets} }
\label{3elint} 
\end{figure}

For $J=3$, one must consider all three-element intervals ${(e_i^{(k)},e_j^{(k')})}$ 
are the causal sets in 
Fig.\ \ref{3elint}. 

Of these, the chain (a) is forbidden, since it won't fit in the interior of a four-layer set.  The others
can all be embedded in layers $\mL_2$ and $\mL_3$, and hence are allowed.  Sets (b) and (d) each
have one embedding: ${\vec n} = (n_2=1,n_3=2)$ for (b) and ${\vec n} = (n_2=2,n_3=1)$ for (d).
Set (c) has two possible embeddings, ${\vec n} = (n_2=1,n_3=2)$ and ${\vec n} = (n_2=2,n_3=1)$.
Set (e), the antichain $\achain_3$,  has four possible embeddings, ${\vec n} = (n_2=2,n_3=1)$,   
${\vec n} = (n_2=1,n_3=2)$,  ${\vec n} = (n_2=3,n_3=0)$, and ${\vec n} = (n_2=0,n_3=3)$.
In all but the last two cases, the initial element $\eki$ must lie in $\mL_1$ and
the final element $\ekpj$ must lie in $\mL_4$.  For the antichain embedding ${\vec n} = (n_2=3,n_3=0)$,
though, $\ekpj$ can lie in either $\mL_3$ or $\mL_4$, and for ${\vec n} = (n_2=0,n_3=3)$, $\eki$ can lie in either
$\mL_1$ or $\mL_2$.  We will see that the ``tight'' intervals dominate.

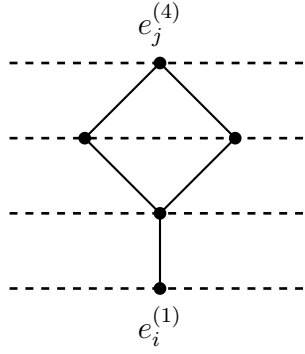
\begin{figure}[!bhtp]
 \begin{center}
   \begin{tikzpicture}[style=thick]
     \draw[line width=1pt, dashed] (-2,0) -- (2,0);
  \draw[line width=1pt, dashed] (-2,1) -- (2,1);
  \draw[line width=1pt, dashed] (-2,2) -- (2,2);
  \draw[line width=1pt, dashed] (-2,3) -- (2,3);
  \draw[color=black] (0,0) -- (0,1);
  \draw[color=black] (0,1) -- (1,2);
  \draw [color=black](0,1) -- (-1,2);
  \draw [color=black](1,2) -- (0,3);
  \draw [color=black](-1,2) -- (0,3); 
 \draw (0,0) node[label=below:$e_i^{(1)}$]   {};
  \draw (0,1) node {};
  \draw (1,2) node {};
  \draw (-1,2) node {};
  \draw (0,3) node[label=above:$e_j^{(4)}$] {};
  \end{tikzpicture}
  \caption{For (b) there are exactly two $4$-element  paths from $e_i^{(1)}$ to $ e_j^{(4)}$, which coincide in the first leg from $e_i^{(1)}$ 
    to an element in $\mL_2$ and then separate thereafter to rejoin at  $ e_j^{(4)}$. }\label{diagb.fig} 
  \end{center}
\end{figure}

Since sets  (b) and (d) are time reversals of each other and have only one realisation each, we 
start with them.  For (b) (see Fig.\ \ref{diagb.fig}), we must require that
  \begin{itemize}
 \item There are two $4$-element paths from $e_i^{(1)} \in \mL_1$ to $e_j^{(4)}\in \mL_4$ such that  there  is only one link
   from $e_i^{(1)} $ to an element  $e_{i'}^{(2)}\in\mL_2$, and subsequently exactly two $3$-element paths  from $e_{i'}^{(2)}$ to 
   $e_j^{(4)}$ via two distinct elements in $\mL_3$.  Such a pattern occurs with probability $p^{(12)}(p^{(23)}p^{(34)})^2=\pi_{1234}\pi_{24}$,
 and comes with combinatorial factors $\gamma_2n$ (the choice of an element in $\mL_2$) and $\binom{\gamma_3n}{2}$
 (the choice of two elements in $\mL_3$).
 \item There are no additional $3$-element paths from $e_i^{(1)} $ to $e_j^{(4)}$ either via an element in $\mL_2$ or an element  
in $\mL_3$.  These give a factor $(1-\pi_{124})^{\gamma_2n-1}(1-\pi_{134})^{\gamma_3n-2}$.
\item There are  no additional $4$-element paths from  $e_i^{(1)} $ to $e_j^{(4)}$. 
This contributes a factor $(1-\pi_{1234})^{\gamma_2\gamma_3n^2 - 2}$.
\end{itemize}
The leading behavior comes from the last term, which has an asymptotic form 
\begin{equation}
q^{(b)} \sim 2^{-\gamma_2\gamma_3n^2|\log_2\alpha_{1234}|}
\label{asympb.eq}
\end{equation}
where, as usual, $\alpha_{1234} = 1-\pi_{1234}$.  Set (d) is the time reversal of set (b), whose
probability can be obtained by simply interchanging layers $\mL_2$ and $\mL_3$.  The leading large $n$
behavior is thus identical to (\ref{asympb.eq}).

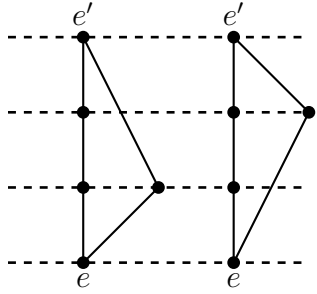
\begin{figure}[!bhtp]
 \begin{center}
   \begin{tikzpicture}[style=thick]
     \draw[line width=1pt, dashed] (-2,0) -- (2,0);
  \draw[line width=1pt, dashed] (-2,1) -- (2,1);
  \draw[line width=1pt, dashed] (-2,2) -- (2,2);
  \draw[line width=1pt, dashed] (-2,3) -- (2,3);
  \draw[color=black] (-1,0) -- (-1,1);
 \draw[color=black] (-1,0) -- (0,1);
  \draw[color=black] (-1,1) -- (-1,2);
  \draw [color=black](-1,2) -- (-1,3);
  \draw [color=black](0,1) -- (-1,3);
\draw (-1,0) node[label=below:$e$]   {};
  \draw (-1,1) node {};
  \draw (-1,2) node {};
  \draw (0,1) node {};
  \draw (-1,3) node[label=above:$e'$] {};

  \draw[color=black] (1,0) -- (1,1);
 \draw[color=black] (1,0) -- (2,2);
  \draw[color=black] (1,1) -- (1,2);
  \draw [color=black](1,2) -- (1,3);
  \draw [color=black](2,2) -- (1,3);
  \draw (1,0) node[label=below:$e$]   {};
  \draw (1,1) node {};
  \draw (1,2) node {};
  \draw (2,2) node {};
  \draw (1,3) node[label=above:$e'$] {};
  \end{tikzpicture}
  \caption{For (c) there are two possible embeddings corresponding to $\vec n_1 = (n_2=2,n_3=1)$ and $\vec n_2 =
    (n_2=1,n_3=2)$.  For $\vec n_1$, there is exactly one $4$-element  path  from $e_i^{(1)} $ to $e_j^{(4)}$ and exactly one
    $3$-element path from $e_i^{(1)} $ to $e_j^{(4)}$ via an element in $\mL_2$. For   $\vec n_2$, there is again exactly one
    $4$-element  path  from $e_i^{(1)} $ to $e_j^{(4)}$ and also exactly one
    $3$-element path from $e_i^{(1)} $ to $e_j^{(4)}$, but now via an element in $\mL_3$.}
  \label{vdiagd.fig} 
  \end{center}
\end{figure}

We next look at set (c) (Fig.\ \ref{vdiagd.fig}), with embeddings $\vec n_1 = (n_2=2,n_3=1)$ and $\vec n_2 =
    (n_2=1,n_3=2)$.
We start with the embedding ${\vec n}_1 = (n_2=2,n_3=1)$. We must now require that
\begin{itemize}
 \item There is one $4$-element  path from $e_i^{(1)} \in \mL_1$ to $e_j^{(4)}\in \mL_4$ via some $e_{i_1}^{(2)} \in \mL_2$ and $e_{j_1}^{(3)} \in \mL_3$.  Such a pattern occurs with probability $\pi_{1234}$,  and comes with combinatorial factors 
 $\gamma_2n$ and $\gamma_3n$. 
 \item There is also  one $3$-element path  from  $e_i^{(1)}$ to $e_j^{(4)}$  via an  element  $e_{i_2}^{(2)}\in\mL_2$ that does not
   intersect the first path at $\mL_2$.  Such a
 pattern occurs with probability $\pi_{124}$,  and comes with a combinatorial factor $\gamma_2n-1$  
(since $e_{i_2}^{(2)}$ must be distinct from the element $e_{i_1}^{(2)}$ in the $4$-element path).
\item There are no additional $3$-element paths from $e_i^{(1)}$ to $e_j^{(4)}$.  Such paths can occur either via elements in
  $\mL_2$ or $\mL_3$ that are not already in either the $4$-element path or the $3$-element path in Fig.\ \ref{vdiagd.fig}.
  This  contributes a 
probability $(1-\pi_{124})^{\gamma_2n-2}(1-\pi_{134})^{\gamma_3n-1}$.  
\item There are no  additional $4$-element paths from  $e_i^{(1)}$ to $e_j^{(4)}$.  
Such an extra path could include either of the elements
$e_{i_1}^{(2)}$ or  $e_{j_1}^{(3)}$, but not both since that would be
a duplicate.  It cannot include $e_{i_2}^{(2)}$ which by assumption is
linked to $e_j^{(4)}$. The resulting factor is thus 
$(1-\pi_{1234})^{\gamma_2\gamma_3n^2 -\gamma_3n- 1}$.
\end{itemize}
We see again that the leading behavior comes from the last term, and that the large $n$ asymptotics
are again given by (\ref{asympb.eq}).  Since the other embedding of set (c), ${\vec n}_2 = (n_2=2,n_3=1)$,
is again a time reversal, its probability can be obtained by interchanging layers $\mL_2$ and $\mL_3$,
and the asymptotic behavior is the same.

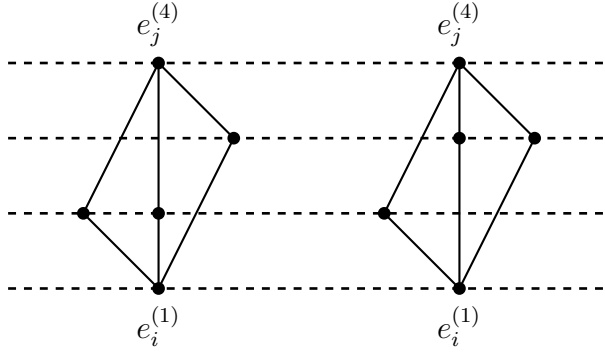
\begin{figure}[!bhtp]
 \begin{center}
   \begin{tikzpicture}[style=thick]
     \draw[line width=1pt, dashed] (-4,0) -- (4,0);
  \draw[line width=1pt, dashed] (-4,1) -- (4,1);
  \draw[line width=1pt, dashed] (-4,2) -- (4,2);
  \draw[line width=1pt, dashed] (-4,3) -- (4,3);
  \draw[color=black] (-2,0) -- (-3,1);
 \draw[color=black] (-2,0) -- (-2,1);
  \draw[color=black] (-2,0) -- (-1,2);
  \draw [color=black](-3,1) -- (-2,3);
  \draw [color=black](-2,1) -- (-2,3);
  \draw [color=black](-1,2) -- (-2,3);
  \draw (-2,0) node[label=below:$e_i^{(1)}$] {};
  \draw (-3,1) node {};
  \draw (-2,1) node {};
  \draw (-1,2) node {}; 
\draw (-2,3) node[label=above:$e_j^{(4)}$] {};

 \draw[color=black] (2,0) -- (1,1);
 \draw[color=black] (2,0) -- (2,2);
  \draw[color=black] (2,0) -- (3,2);
  \draw [color=black](1,1) -- (2,3);
  \draw [color=black](2,2) -- (2,3);
  \draw [color=black](3,2) -- (2,3);
  \draw (2,0) node[label=below: $e_i^{(1)}$] {};
  \draw (1,1) node {};
  \draw (2,2) node {};
  \draw (3,2) node {}; 
  \draw (2,3) node[label=above: $e_j^{(4)}$] {};
  \end{tikzpicture}
  \caption{For (e) with $e_i^{(1)} \in \mL_1$, and $e_j^{(4)}\in \mL_4$ there are two possible embeddings,
   corresponding to $\vec n_1 = (n_2=2,n_3=1)$ and $\vec n_2 =
    (n_2=1,n_3=2)$.  For each there are exactly  three $3$-element paths from $e_i^{(1)}$, to $e_j^{(4)}$ .
    In the former, two paths pass
    through $\mL_2$ and  the third  through $\mL_3$, while in the latter,  one path passes 
    through  $\mL_2$ and  the other two through $\mL_3$.}
  \label{vdiage.fig} 
  \end{center}
\end{figure}
Finally, we turn to the antichain, set (e).
Begin with the embedding ${\vec n}_1 = (n_2=2,n_3=1)$ (Fig.\ \ref{vdiage.fig}).
We must now require that
\begin{itemize}
 \item There are exactly two $3$-element paths (and no more) from $e_i^{(1)} \in \mL_1$ to $e_j^{(4)}\in \mL_4$ via two elements $e_{i_1}^{(2)},e_{i_2}^{(2)}\in\mL_2$.   These give a contribution $(\pi_{124})^2(1-\pi_{124})^{\gamma_2n-2}$
 to the probability, with a  combinatorial  factor $\binom{\gamma_2n}{2}$.
 \item There is exactly one path $3$-element path from $e_i^{(1)}$ to an element $e_{j_1}^{(3)}\in\mL_3$ to $e_j^{(4)}$.  This gives
 a contribution $\pi_{134}(1- \pi_{134})^{\gamma_3n-1}$, with a combinatorial factor $\gamma_3n$.
\item There is no $4$-element path from $e_i^{(1)}$ to $e_j^{(4)}$.  {Such an extra path cannot include the elements
  $e_{i_1}^{(2)}$ or $e_{i_2}^{(2)}$ in $\mL_2$, since by assumption these are links---that is,
nearest neighbors---to $e_j^{(4)}$.  Similarly, it cannot contain $e_{j_1}^{(3)}$, which is linked to $e_i^{(1)}$.}  The resulting
factor is thus $(1-\pi_{1234})^{(\gamma_2n-2)(\gamma_3n - 1)}$.
\end{itemize}
As this example makes clear, the fine details of the combinatorics can be complicated.  But the leading
asymptotic behavior is not affected; it is again given by (\ref{asympb.eq}).  The embedding 
${\vec n}_2= (n_2=1,n_3=2)$ is again related by time reversal, and has the same asymptotic behavior.

\begin{figure}[!bhtp]
 \begin{center}
   \begin{tikzpicture}[style=thick]
     \draw[line width=1pt, dashed] (-4,0) -- (4,0);
  \draw[line width=1pt, dashed] (-4,1) -- (4,1);
  \draw[line width=1pt, dashed] (-4,2) -- (4,2);
  \draw[line width=1pt, dashed] (-4,3) -- (4,3);
  \draw[color=black] (-2,0) -- (-3,1);
 \draw[color=black] (-2,0) -- (-2,1);
  \draw[color=black] (-2,0) -- (-1,1);
  \draw [color=black](-3,1) -- (-2,3);
  \draw [color=black](-2,1) -- (-2,3);
  \draw [color=black](-1,1) -- (-2,3);
  \draw (-2,0) node[label=below:$e_i^{(1)}$] {};
  \draw (-3,1) node {};
  \draw (-2,1) node {};
  \draw (-1,1) node {}; 
\draw (-2,3) node[label=above:$e_j^{(4)}$] {};

 \draw[color=black] (2,0) -- (1,2);
 \draw[color=black] (2,0) -- (2,2);
  \draw[color=black] (2,0) -- (3,2);
  \draw [color=black](1,2) -- (2,3);
  \draw [color=black](2,2) -- (2,3);
  \draw [color=black](3,2) -- (2,3);
  \draw (2,0) node[label=below:$e_i^{(1)}$] {};
  \draw (1,2) node {};
  \draw (2,2) node {};
  \draw (3,2) node {}; 
  \draw (2,3) node[label=above:$e_j^{(4)}$] {};
  \end{tikzpicture}
  \caption{For 
     (e) with $e_i^{(1)} \in \mL_1$, and $e_j^{(4)}\in \mL_4$ there are two other possible embeddings,
   corresponding to $\vec n_3 = (n_2=3,n_3=0)$ and $\vec n_4 =
    (n_2=0,n_3=3)$.  For each there are exactly  three $3$-element paths from $e_i^{(1)} $ to $e_j^{(4)}$; in the former, all pass 
    through $\mL_2$, while  in the latter,  all  pass
    through $\mL_3$.}
  \label{vdiagc.fig} 
  \end{center}
\end{figure}
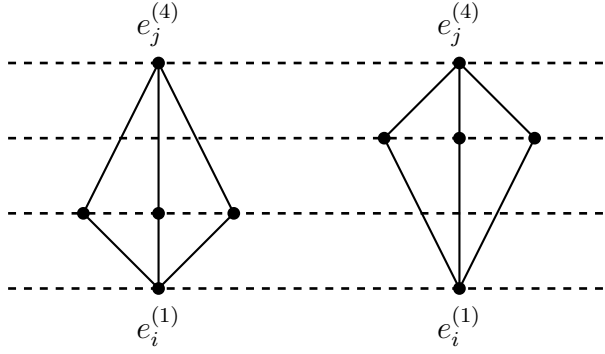

Next, we consider the antichain embedding ${\vec n}_3 = (n_2=3,n_3=0)$ (Fig.\ \ref{vdiagc.fig}).  There are now two possible
configurations: the final element $e_j^{(k)}$ can lie in layer three  (i.e., $(k,k')=(1,3)$) or in layer four (i.e., $(k,k')=(1,4)$).

We start with $(k,k')=(1,4)$. We must require that
\begin{itemize}
 \item There are exactly three $3$-element  paths from $e_i^{(1)} \in \mL_1$ to $e_j^{(4)}\in \mL_4$ via  elements
   $e_{i_1}^{(2)}, e_{i_2}^{(2)} , e_{i_3}^{(2)}  \in\mL_2$.   This
   gives a contribution $(\pi_{124})^3(1-\pi_{124})^{\gamma_2n-3}$
 to the probability, with a  combinatorial  factor $\binom{\gamma_2n}{3}$.
 \item There is no $3$-element path from  $e_i^{(1)}$ to $e_j^{(4)}$ via elements in $\mL_3$.  This gives  a contribution 
 $(1-\pi_{134})^{\gamma_3n}$, with a combinatorial factor $\gamma_3n$.
\item There is no additional $4$-element path from $e_i^{(1)}$ to $e_j^{(4)}$.  
As in the previous example, such an extra path cannot include elements  $e_{i_1}^{(2)}, e_{i_2}^{(2)} $ or $
e_{i_3}^{(2)} $. 
The resulting factor is thus $(1-\pi_{1234})^{(\gamma_2n-3)\gamma_3n}$.
\end{itemize}
Once again, the last term dominates the asymptotic behavior, which is again given by (\ref{asympb.eq}).
As before, the embedding ${\vec n}_4 = (n_2=0,n_3=3)$ with $(k,k')=(1,4)$ is the time reversal, and has the 
same asymptotic behavior.

We now have two remaining possibilities, the ``tight fit'' cases with $k'=k+2$. There are two possibilities here:
$(k,k')=(1,3)$ with  ${\vec n} = (n_2=3,n_3=0)$ 
and $(k,k')=(2,4)$ with ${\vec n} = (n_2=0,n_3=3)$.  But these are already 
familiar from the $K=3$-layer case.  For $(k,k')=(1,3)$, layer $\mL_4$ is irrelevant; all that matters is the three-layer
set $\mL_1\cup \mL_2\cup \mL_3$, for which the computation was already done in section \ref{secK3.ssec},
with a probability given by (\ref{q4.eq}).  Similarly, for $(k,k')=(2,4)$, layer $\mL_1$ is irrelevant, and
the result is again given by (\ref{q4.eq}).

Comparing (\ref{q4.eq}) and (\ref{asympb.eq}), we see that at large $n$, the probabilities for tight fit 
antichains fall off  exponentially more slowly than any of the other intervals. 
   By the argument 
of section \ref{strategy.ssec}, these also dominate the probabilities $\cP(N_3=R)$, reducing the
problem to the three-layer case.

\subsection{General $K,J$ \label{mainKJ.ssec}}

The experience with four-layer sets with $J=1,2,3$ suggests  that the dominant contribution to 
$\cP(N_J=R)$ comes from tight fit antichains $\achain_J$. As a first step, we therefore calculate 
$\cP(N_J^{\achain_J})$ and compare the contributions from the tight fit and loose fit antichains for 
general $K,J$. We find that, as expected, the former dominate. 

\subsubsection{{Antichain Contributions} \label{antichains.sssec}}

For $J>0$, the  allowed $(k,k')$ pairs are  $k=1, \ldots, K-2$ and $k'=k+2, \ldots K $.
We begin with the by now familiar calculation of the contribution from tight fit antichains, that is, antichains with  $k'=k+2$. 
As in the three-layer case, these admit a single embedding, ${\vec n} =  (\dots,n_k=0, n_{k+1} =J, n_{k+2}=0, \dots)$. 
The probability for 
the interval $(\eki,e_j^{(k+2)})$ to be the antichain $\achain_J$ can be simply copied from 
the $K=3$ case of section \ref{secK3.ssec}:
\begin{equation}
q_k= \binom{\ff_{k+1} n}{J} (\pi_k)^J( 1-\pi_k)^{\ff_{k+1}n-J}  \, \, \quad \hbox{with $\pi_k \equiv p^{(k(k+1))}  p^{((k+1)(k+2))}$} ,
  \end{equation} 
a trivial generalisation of Eqn.\ \eqref{q3.eq}.  Thus for fixed $J$, to leading order in $n$,
\begin{equation}
  q_k \sim  2^{-\ff_{k+1} n |\log_2 (\alpha_k)|}  
  \label{antichain.eq} 
  \end{equation} 
with $\alpha_k = 1-\pi_k$.

Next consider loose fit antichains, that is, antichains with $k'>k+2$.  Suppose there are $r>1$ distinct layers sandwiched between $k$ and $k'$,
that is, $k'=k+r+1$. 
For a fixed $(k,k')$ the distribution of elements is again a weak composition problem: we wish to distribute $J$
elements among $r$ layers.  There are  $\binom{r+J-1}{J-1}$ such embeddings.  Consider one such embedding 
\begin{equation}
{\vec n}=(\dots,n_{k_1}, n_{k_2} \ldots n_{k_s},\dots), \ \hbox{with $k<k_1 \ldots < k_s < k'$, $1 < s \leq r$,  and}\ \sum_{i=1}^s n_{k_i}=J .
\label{embed1.eq}
\end{equation} 
Fix an initial element $\eki \in\mL_k$ and a final element $\ekpj \in\mL_{k'}$.  In order for the interval $(\eki,\ekpj)$ to be the antichain 
$\achain_J$, we must require
\begin{itemize}
\item For each $k_i$, there are exactly $n_{k_i}$ distinct $3$-element  paths from $\eki$ to $\ekpj$ via an element  in layer $\mL_{k_i}$.  Each $k_i$
 thus contributes a factor 
 $$\binom{\gamma_{k_i}n}{n_{k_i}}\pi_{k_i}^{n_{k_i}}(1-\pi_{k_i})^{\gamma_{k_i}n-n_{k_i}} ,$$
 where $\pi_{k_i}=p^{(kk_i)}p^{(k_ik')}$.
\item  There are no $t$-element paths from $\eki$ to  $\ekpj$ for $t >3$. 
 In terms of the link matrix, this is  equivalent 
to the requirement that $(\ccL^{t-1})_{ij}=0$ for every $3<  t \le r+2 $. 
 \end{itemize}
Thus
\begin{equation}
 q_{kk'}=\biggl(\prod_{i=1}^{s} \binom{\gamma_{k_i}n}{n_{k_i}}
\pi_{k_i}^{n_{k_i}}(1-\pi_{k_i})^{\gamma_{k_i}n-n_{k_i}}\biggr) \, \,  \prod_{t=4}^{r+2}\cP\left((\ccL^{t-1})_{ij}=0\right) 
\end{equation} 

The suppression for $r>1$ comes from the second product: multi-element paths from $\eki$ to $\ekpj$ are common, and
intervals that exclude them are very rare.  The dominant contribution comes from {$t=r+2$}, for which
 \begin{equation}
          \cP((\ccL^{r+1})_{ij}=0) = (1-\pi_{{r}})^{\Gamma_{r}n^r {+ \mathcal{O}(n^{r-1})}}  ,
 \label{suppress.eq}
 \end{equation} 
where $\pi_{r}=p^{(k(k+1))}p^{((k+1)(k+2))} \ldots  p^{((k+r-1)(k+r))}p^{((k+r)(k+r+1))} $ and $\Gamma_r = \gamma_{k+1}
\gamma_{k+2} \ldots \gamma_{k+r}$.  {As in section \ref{J3.sec}, there are corrections to the exponent coming from 
the fact that a small set of multi-element paths are already excluded.  If we know that the element $e_1\in\mL_{k_1}$  is a
member of our candidate antichain (\ref{embed1.eq}),  that is, $e\link e_1\link e'$ with
{$e_1\in\achain_J$}}, 
then by the definition of a link, there can be no multi-element
paths of the form $e\link e_1 \link e_2\link e'$ or $e\link e_2\link
e_1\link e'$. {But the resulting modifications  are subdominant; at most
they require that we replace some factors $\gamma_{k_i}n$ in the exponent with $\gamma_{k_i}(n-n_{k_i})$,
with $n_{k_i}\leq J\ll n$}.
Thus 
\begin{equation}
q_{kk'} \sim  2^{-\Gamma_r n^r |\log_2\alpha_r|} ,
  \end{equation} 
where again $\alpha_r = 1-\pi_r$.  Comparing to Eqn.\ (\ref{antichain.eq}), we see that the contribution of the loose
antichains $(r>1)$ is subdominant.  Hence the leading order contribution to $\cP(N_J^{\achain_J}=R)$ comes 
 from  the tight antichains, with
 \begin{equation}
 \cP(N_J^{\achain_J}=R) \sim 2^{-{\tilde\gamma}|\log_2{\tilde\alpha}| R n } ,
 \label{Pachain.eq}
 \end{equation} 
where ${\tilde\gamma}|\log_2{\tilde\alpha}|$ is the smallest of the $\ff_{k+1} |\log_2\alpha_k|$ for $k \in (1, \ldots K-2)$.  

\subsubsection{{Non-Antichain Contributions}}

We will now prove that for any allowed $c \in \Omega_J$ that is not the antichain, and for any pair of layers $(k,k')$, the
probability that an interval $(\eki,\ekpj)$ is 
 $c$
with embedding $\vec n$ is, to leading order,  
\begin{equation}
 q(c)\sim 2^{-\eta(c) n^a}
\label{result.eq} 
\end{equation} 
where $\eta(c) > 0$ and $a>1$.  This implies that for large $n$, the tight antichains of the preceding subsection are
always dominant.

{First note that any causal set $c$ that is not an antichain has a height $h(c) >1$, as defined in section
\ref{prelim.sec}. Hence unless  $k'-k =r+1 \geq h(c) +1$, such a $c$ cannot contribute, as we saw 
  already for the $3$-element chain in the $K=4, J=3$ calculation of section \ref{J3.sec}. Therefore we need $r \geq h(c) > 1 $,
 which also means that $K-2 \geq h(c) >1$.} 

Consider a causal set $c$ with an embedding labelled by $(k,k',{\vec n} = (\dots,n_{k_1},n_{k_2},\dots,n_{k_s},\dots))$,
 with $k<k_1 \ldots < k_s < k'$ and $\sum_{i=1}^s n_{k_i}=J$, {$1 < h(c) \leq s \leq r $}.  The probability $q(c;k,k',{\vec n})$ will have two
contributions:
\begin{itemize}
\item The set $c$ must occur, with the embedding $(k,k',{\vec n})$, 
 as the interval $(\eki,\ekpj)$.  Call this probability $\cP_1$.
\item There must be no other paths connecting $\eki$ and $\ekpj$.  Call this probability $\cP_2$.
\end{itemize}

We first place an upper bound on $\cP_1$.  This probability will depend on the topology of $c$, and will in
general be quite complicated.  Given an exact embedding $c\hookrightarrow (\eki,\ekpj)$, though, this probability
is certainly bounded by $1$.  

Note, however, that a specification of $\vec n$ does not completely determine the 
embedding of $c$, since we are still free to choose the locations of $n_{k_1}$ elements in layer $\mL_{k_1}$, $n_{k_2}$ 
elements in layer $\mL_{k_2}$, etc.  This gives an additional combinatorial factor of
\begin{equation}
\binom{\gamma_{k_1}n}{n_{k_1}}\binom{\gamma_{k_2}n}{n_{k_2}}\dots\binom{\gamma_{k_s}n}{n_{k_s}}
   \sim n^{n_{k_1}}n^{n_{k_2}}\dots n^{n_{k_s}} = n^J
\end{equation}
where the asymptotic behavior comes from the fact that $\gamma_{k_i}n \gg J > n_{k_i}$.  Hence $\cP_1$ is bounded
above by a constant times $n^J$.  Note that this matches the behavior we found for $J=1,2,3$  in the $K=4$ layer
case of sections \ref{J1.sec}{--\ref{J3.sec}.} 

We now turn to $\cP_2$, the probability that there is no other path from $\eki$ to $\ekpj$.  {For any finite $J$ element
  interval, or equivalently any finite $c$, of cardinality $J \ll n$,  there are only a finite number of
paths from $\eki$ to $\ekpj$ in $c$.  All others in $C$ must be excluded. } Therefore the argument is virtually
the same as that of section \ref{antichains.sssec}.  As in that case, the leading suppression comes from the need
to exclude any additional {$(k'-k+1)$-element paths} that start at $\eki \in\mL_k$, hit all of the intermediate layers 
$\mL_{k+1},\mL_{k+2},\dots,\mL_{k'-1}$, and end at $\ekpj \in\mL_{k'}$.   

The one new complication comes from
the word ``additional'': there may be {$(k'-k+1)$-element} paths that {already} lie in $c$, and {others that
intersect with $c$ in a way that already excludes them, like the paths described after Eqn.\ \ref{suppress.eq}.  These}
should not be {separately} excluded.  We can
again obtain an upper bound, though, by considering paths that do not intersect $c$ at all.  That is,  we 
 must at least 
exclude any path that starts at $\eki\in \mL_k$, links to any of the $\gamma_{k+1}n-n_{k+1}$ elements in layer 
$\mL_{k+1}$ that are not in $c$, then to any of the $\gamma_{k+2}n-n_{k+2}$ elements in layer $\mL_{k+2}$ that are 
not in $c$, and so on, and ends at $\ekpj \in \mL_{k'}$. For large $n$, the number of such paths is
\begin{equation}
(\gamma_{k+1}n-n_{k+1})(\gamma_{k+2}n-n_{k+2})\dots(\gamma_{k'-1}n-n_{k'}) \sim \Gamma_rn^r {+ \mathcal{O}(n^{r-1})},
\end{equation}
where, again, $r=k'-k-1$ and $\Gamma_r = \gamma_{k+1}\gamma_{k+2} \ldots \gamma_{k+r}$,
just as in Eqn.\ (\ref{suppress.eq}).  Thus, asymptotically,
\begin{equation}
q(c;k,k',{\vec n}) = \cP_1\cP_2 \lesssim 2^{-\Gamma_r n^r |\log_2\alpha_r|} ,
\label{csuppress.eq}
\end{equation} 
where again $\alpha_r = 1-\pi_r$.
 
Comparing to Eqn.\ (\ref{antichain.eq}), we see that the probabilities are dominated by those of tight
fit antichains, whose probabilities take the same form as those for the three-layer sets discussed in
section \ref{secK3.ssec}.  One might worry that an ``entropy'' factor from the number of causal sets $c$ 
might overwhelm the suppression (\ref{csuppress.eq}).   But the number of $J$-element causal sets grows as 
$2^{J^2/4}$, so as long as $n\gg J$,  this will not happen.  An additional combinatorial factor can come from
the number of weak partitions, that is, the number of ways of
distributing $N_J$ over different intervals. 
 As discussed in section \ref{strategy.ssec}, however, these are also not large enough to overcome the
exponential suppression (\ref{csuppress.eq}).

\section{Conclusions \label{conclusions.sec}}

Let us summarize our results and place them in context.  Our interest is in causal set theory as a discrete
model of spacetime.  The fundamental problem we are addressing is the fact that almost all causal sets
are not, in fact, anything like spacetimes.  The set of causal sets is dominated by ``layered'' sets, sets
with only a small number of moments of time, which must somehow be suppressed in any physically
sensible model.

An obvious candidate for such a physically sensible model is the path integral (technically, the path sum) 
with the BDG action, the discrete version of the Einstein-Hilbert action.  In \cite{lc}, it was shown that
for a wide range of coupling constants, this path integral does, in fact, very strongly suppress the simplest
layered causal sets, the bilayer sets. In \cite{ams} this result was
extended to  {the dominant class of K-layered sets of \cite{dharone,pst}}, but using  the link action, a truncation of the BDG action, which no longer approximates the continuum Einstein-Hilbert action.

In \cite{ccs}, we took the next step, by showing that for three-layer sets this truncation is harmless.
Specifically, in the limit that the number $n$ of causal set elements becomes large, the difference between
the BDG action and the link action becomes
negligible for almost all three-layer sets. What we have proved in this
paper is that the same is true for $K$-layer causal sets, for any $K\ll n$.    {As discussed in the present paper, the counting
  arguments use labelled causal sets, but to leading order in $n$, they
  also hold for unlabelled  causal sets. Hence, in conjunction with
  the result of \cite{ams},  this implies that } the BDG path sum%
---the causal set version of the standard gravitational path integral---strongly suppresses the entropically
dominant layered causal sets. 

It is worth emphasizing that the suppression is \emph{extremely} strong.  For a causal set with $n$ elements,
the suppression factor takes the form $2^{-kn^2}$.  If the discreteness scale is the Planck scale, a spacetime
region of $1\,\hbox{ns}\times 1\,\hbox{cm}^3$ has $n\sim 10^{133}$, giving a suppression of order $2^{-10^{266}}$.

The BDG action is one of many nonlocal actions, which are constructed using a nonlocality scale $\xi \geq \ell$ \cite{bd,dg,glaser}.  As $\xi$
becomes larger compared to the discreteness scale $\ell$, it brings in more
contributions from larger $J$-element intervals to the action. Since our analysis works for all $K,J \ll n$, {though},
these still do not
contribute significantly to the nonlocal action, which therefore {again} reduces to the link action. 

While an important question in causal set theory has been settled by these arguments,
a number of other questions remain.  As noted at the end of section 2.1, there remains a
class of ``sparse'' layered sets---albeit one of measure zero--- for which our analysis
does not apply.  Moreover, of the remaining non-layered causal sets in $\Omega_n$, many,
and  plausibly most, are also not continuumlike.  What role do they play in the
semiclassical regime? Even if we were to restrict ourselves to continuumlike causal
sets, these can have any spacetime dimension. Does an action of dimension $d$ then pick
out only spacetimes of dimension $d$ and suppress those of dimension $d'\ne d$? These are
important questions going forward.

\vspace{1.5ex}
\begin{flushleft}
\large\bf Acknowledgments
\end{flushleft}

S.~C.\ was supported in part by Department of Energy grant
DE-FG02-91ER40674. S.~S.\ was supported in part by the Blaumann
Foundation Grant, Call II as well as the SERB/F/11952/2023-2024DST
Matrics Grant.


\end{document}